\documentclass[]{aero} % <----- add lineno in the square brackets
\usepackage{siunitx}
\usepackage{amsmath}
\usepackage{amssymb}
\usepackage{wasysym}
\usepackage{tikz}
\usepackage{booktabs}  % For professional-looking tables
\usepackage{array}     % For column width control
\usepackage{makecell}  % Allows multi-line cells
\usepackage{caption}
\usepackage{subcaption}
\usepackage{threeparttable}
\usepackage{multirow}
\usepackage[table]{xcolor}
\usepackage{pifont}
\usepackage{nicefrac}
\usepackage{float}
\newcommand{\rev}[1]{#1}
\newcommand{\imgrev}[1]{#1}

\usetikzlibrary{calc,patterns,angles,quotes}

\begin{document}
\title{MasconCube: Fast and Accurate Gravity Modeling with an Explicit Representation}
\author{Pietro Fanti$^{1}$, Dario Izzo$^{1}$\cor{}}
\maketitle  % make title page
\noindent\cor{} dario.izzo@esa.int \\
\\
\noindent\textit{1. Advanced Concepts Team, European Space Research and Technology Centre (ESTEC), Noordwijk, The Netherlands.}
\begin{abstract}
    \normalsize
    The geodesy of irregularly shaped small bodies presents fundamental challenges for gravitational field modeling, particularly as deep space exploration missions increasingly target asteroids and comets. Traditional approaches suffer from critical limitations: spherical harmonics diverge within the Brillouin sphere where spacecraft typically operate, polyhedral models assume unrealistic homogeneous density distributions, and existing machine learning methods like GeodesyNets and Physics-Informed Neural Networks (PINN-GM) require extensive computational resources and training time. This work introduces MasconCubes, a novel self-supervised learning approach that formulates gravity inversion as a direct optimization problem over a regular 3D grid of point masses (mascons). Unlike implicit neural representations, MasconCubes explicitly model mass distributions while leveraging known asteroid shape information to constrain the solution space. Comprehensive evaluation on diverse asteroid models including Bennu, Eros, Itokawa, and synthetic planetesimals demonstrates that MasconCubes achieve superior performance across multiple metrics. Most notably, MasconCubes demonstrate computational efficiency advantages with training times approximately 40 times faster than GeodesyNets while maintaining physical interpretability through explicit mass distributions. These results establish MasconCubes as a promising approach for mission-critical gravitational modeling applications requiring high accuracy, computational efficiency, and physical insight into internal mass distributions of irregular celestial bodies.
\end{abstract}
\clearpage
% -- Nomenclature

% \Nomenclature
% % \note{\vspace*{-1.4em} (Or others such as Abbreviation, if necessary; nomenclature entries should have the units identified)}

% \begin{center}
% \begin{tabular}{|P{.45\linewidth}|P{.45\linewidth}|}
% \hline
% $u$ & Optical flow in camera x direction \\
% $v$ & Optical flow in camera y direction \\
% $p$ & Angular velocity ($\frac{\text{rad}}{\text{s}}$) \\
% $q$ & Angular velocity ($\frac{\text{rad}}{\text{s}}$) \\
% $r$ & Angular velocity ($\frac{\text{rad}}{\text{s}}$) \\
% $r$ & position (m) \\
% $X$ & x position (m) \\
% $Y$ & y position (m) \\
% $Z$ & z position (m) \\
% $x$ & scaled x position (m) \\
% $y$ & scaled y position (m) \\
% $k$ & unit normal vector \\
% $\alpha$ & x-component of $\hat{k}$ \\
% $\beta$ & y-component of $\hat{k}$ \\
% $\gamma$ & z-component of $\hat{k}$ \\
% $H$ & Altitude (m) \\
% $h$ & Inverse depth map ($\frac{1}{\text{m}}$) \\
% $\epsilon$ & Rotational angle (rad) \\
% $R$ & Radius of the Moon (m) \\

% \hline
% %$a_{\rm c}$ & cylinder diameter (cm) \\
% O & Spacecraft \\
% S & Nadir point on surface of Moon \\
% P & Line of sight point on surface of Moon \\
% M & center of the Moon \\

% \hline
% $\mathbf{R}$ & Rotation matrix \\
% $\mathbf{\hat{x}}$ & Reference frame x-axis \\
% $\vec{\square}^c$ & Vector expressed in c reference frame \\
% $\vec{\square}_{AB}$ & Vector from A to B \\

% \hline
% \end{tabular}
% \end{center}

\section{Introduction}

The geodesy of irregularly shaped small bodies, particularly asteroids and comets, presents fundamental challenges that have driven decades of research in gravitational field modeling. As deep space exploration missions increasingly target these celestial objects, from previous missions like NEAR-Shoemaker~\cite{prockter2002near, veverka2000near} and Hayabusa~\cite{kawaguchi2008hayabusa, yano2006hayabusa, fujiwara2006hayabusa} to current endeavors such as Hera~\cite{esa_hera_2024}, Psyche~\cite{nasa_psyche_2023}, and Tianwen-2~\cite{zhang2025system}, accurate characterization of their internal mass distributions has become essential for mission planning, proximity operations, and scientific understanding. \rev{ Moreover, precise gravity models are essential for studying the evolution of natural systems, where higher-order harmonics can influence features such as dust ring formation~\cite{liu2021configuration}.}

The problem of inferring internal density distributions from external gravity measurements is inherently ill-posed, presenting significant computational and mathematical challenges~\cite{blakely1996potential}. Unlike large planetary bodies with relatively regular shapes, asteroids and comets exhibit highly irregular geometries that complicate traditional gravitational modeling approaches. The sparse and noisy nature of gravity measurements, combined with the non-unique solutions inherent to gravity inversion, makes the reconstruction of internal mass distributions particularly challenging. Furthermore, the computational complexity~\cite{martin2025physics} required to achieve high-fidelity models while maintaining numerical stability near these irregular surfaces compounds these difficulties.

Spherical harmonics have long served as the foundation for gravitational field representation~\cite{brillouin1933equations, hashimoto2010vision, accomazzo2017final}, offering analytical solutions to Laplace's equation and providing a mathematically elegant framework for potential field description. However, their application to irregular small bodies faces critical spectral limitations~\cite{scheeres2000estimating, hirt2017convergence, sebera2016spheroidal}.

Polyhedral gravity models~\cite{tsoulis2012analytical, paul1974gravity, d2014analytical} have emerged as a popular alternative, providing analytical solutions that remain stable within the Brillouin sphere and down to the body's surface. However, they carry a fundamental limitation: they assume a homogeneous density distribution throughout the body. This assumption may not accurately reflect the internal composition of real asteroids, which can exhibit significant density variations due to porosity, composition changes, or internal voids.

\rev{Recent work addresses gravity field estimation through both observational and computational approaches. Concepts like SmallSat swarm gravimetry~\cite{ledbetter2021smallsat} propose distributed measurements to infer bulk gravitational parameters, while mascon-based models~\cite{shang2017efficient} use discretized polyhedral sources for orbital dynamics analyses. These efforts typically emphasize measurement strategies or forward gravitational computation rather than direct inference of internal structure.}

% Recent advances in machine learning have introduced GeodesyNets~\cite{izzo2022geodesy, schuhmacher2023geodesy}, neural networks that learn three-dimensional density distributions as continuous functions. These networks represent the density field as a neural implicit function, allowing for both shape reconstruction and internal structure inference without requiring prior shape knowledge. GeodesyNets demonstrate remarkable capability in capturing complex gravitational fields with relative errors below $1\%$ even near asteroid surfaces. However, the training process can be computationally intensive, particularly for complex heterogeneous models, and requires several data points.

Recent advances in machine learning have introduced \rev{various neural network architectures for asteroid gravity modeling. Early approaches explored Extreme Learning Machines (ELMs)~\cite{furfaro2021elm}, which employ single-layer feedforward networks trained without iterative tuning to learn gravitational acceleration around irregular bodies, demonstrating fast training times suitable for on-board computation. Hopfield Neural Networks have been investigated~\cite{pasquale2022small, zhao2023board} for on-board reconstruction of spherical harmonics coefficients, showing comparable performance to Extended Kalman Filtering approaches for parameter estimation with reduced computational requirements. Concurrently, GeodesyNets~\cite{izzo2022geodesy, schuhmacher2023geodesy} have emerged as neural implicit representations} that learn three-dimensional density distributions as continuous functions\rev{,} allowing for both shape reconstruction and internal structure inference without requiring prior shape knowledge\rev{, demonstrating} remarkable capability in capturing complex gravitational fields with relative errors below $1\%$ even near asteroid surfaces. Physics-Informed Neural Networks for Gravity Modeling (PINN-GM)~\cite{martin2022pinn1, martin2022pinn2}, with their latest iteration PINN-GM III~\cite{martin2025pinn}, represent another machine learning approach that incorporates physical constraints directly into neural network training to learn gravitational potential. These models demonstrate advantages in compactness, noise robustness, and sample efficiency, but remain computationally expensive to train.

% Physics-Informed Neural Networks for Gravity Modeling (PINN-GM)~\cite{martin2022pinn1, martin2022pinn2}, with their latest iteration PINN-GM III~\cite{martin2025pinn}, represent another machine learning approach that incorporates physical constraints directly into neural network training to learn gravitational potential. These models demonstrate advantages in compactness, noise robustness, and sample efficiency, but remain computationally expensive to train. Although PINN-GM shows promise for gravity modeling, their capability for geodesy reconstruction has been speculated, but never empirically tested.

Current approaches exhibit several critical limitations that motivate the development of alternative methodologies. Training efficiency remains a persistent challenge across machine learning approaches, with GeodesyNets and PINN-GM models often requiring extensive computational resources and training time to achieve convergence. The spherical harmonics fidelity problem, where traditional analytical methods fail to capture fine-scale gravitational features of irregular bodies, continues to limit mission-critical applications requiring high accuracy near surfaces. Reproducibility concerns arise from the complex hyperparameter spaces and initialization sensitivities of neural network approaches, making it difficult to consistently achieve optimal performance across different types of asteroid and mission scenarios.

Additionally, existing methods often struggle with the trade-off between computational efficiency and physical accuracy. While mascon models offer direct physical interpretation through point mass distributions, they traditionally require large numbers of masses to achieve fine-grained accuracy, leading to computational bottlenecks. The lack of a unified approach that combines the physical interpretability of mascon methods with the representational power of modern machine learning techniques represents a significant gap in current capabilities.

The MasconCube approach introduced in this work addresses these limitations by formulating gravity inversion as a direct optimization problem over a regular 3D grid of point masses, leveraging self-supervised learning to reconstruct internal mass distributions efficiently while maintaining physical interpretability and computational tractability.

\section{Methodology}

This section presents the comprehensive methodology employed to develop, train, and evaluate MasconCubes for asteroid gravity modeling. We begin by describing the diverse suite of simulated asteroid models constructed for benchmarking. We then detail the  training approach used to optimize the mass distribution within the 3D mascon grid, including the observation model, loss function, and optimization procedures. Finally, we outline the multi-faceted evaluation framework that assesses model performance.
 
\subsection{Asteroid Models}
\label{sec:asteroids}

To rigorously evaluate the performance of MasconCubes and benchmark it against GeodesyNets and PINN-GM III, we constructed a suite of simulated asteroid models derived mostly from well-characterized real-world bodies. Our selection criteria emphasized diversity in shape complexity, size, and scientific relevance, ensuring that our experiments capture a representative range of geodetic challenges encountered in planetary science.

We selected a fictitious planetesimal body previously used in the literature \cite{izzo2022geodesy, martin2025pinn}, along with three asteroids for which high-resolution shape models are publicly available: 101955 Bennu, 433 Eros, and 25143 Itokawa. For consistency with prior studies \cite{izzo2022geodesy}, we employed simplified, low-resolution versions of their surface meshes. To operate in a dimensionless framework, \rev{each surface model was uniformly scaled so that its outermost vertex lay on a sphere of radius 0.8}.

\rev{From these surface meshes, the internal volumes were discretized into tetrahedra using the TetGen~\cite{si2006quality} implementation of the Delaunay refinement algorithm~\cite{ruppert1995delaunay,shewchuk1998tetrahedral}. This volumetric meshing step generated a fully connected 3D grid that filled the interior of each asteroid while preserving its surface topology. A mass concentration (mascon) was assigned to the centroid of each tetrahedron, with a dimensionless mass equal to its volume.} The mascon models were subsequently normalized by dividing each mass by the total mass of the body, ensuring that the overall mass sums to 1.

\begin{figure}
    \centering
    \subfigure[Bennu - Two regions]{
        \includegraphics[width=0.31\linewidth]{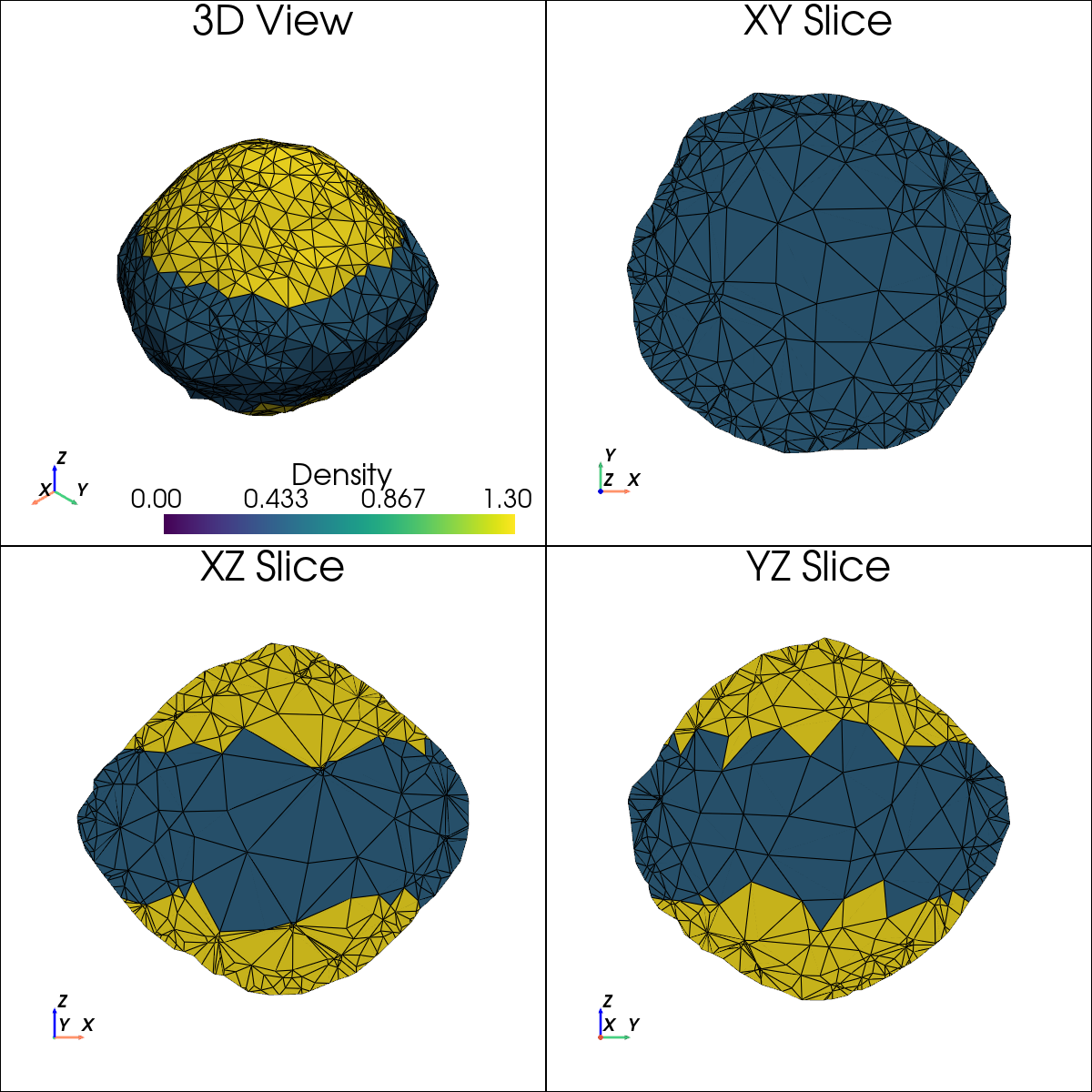}
        \label{fig:mesh_bennu}
    }
    \subfigure[Eros - Two regions]{
        \includegraphics[width=0.31\linewidth]{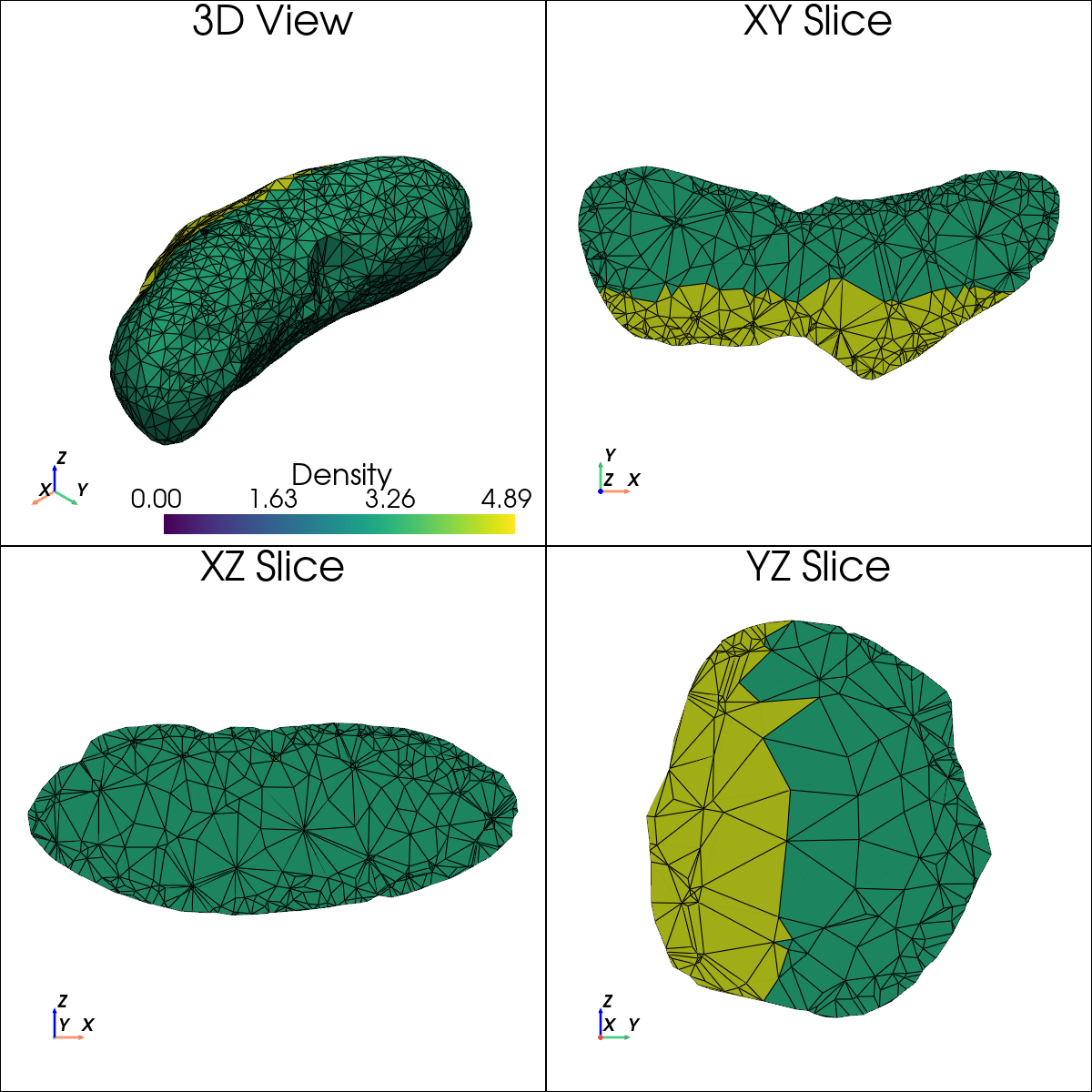}
        \label{fig:mesh_eros_2}
    }
    \subfigure[Eros - Three regions]{
        \includegraphics[width=0.31\linewidth]{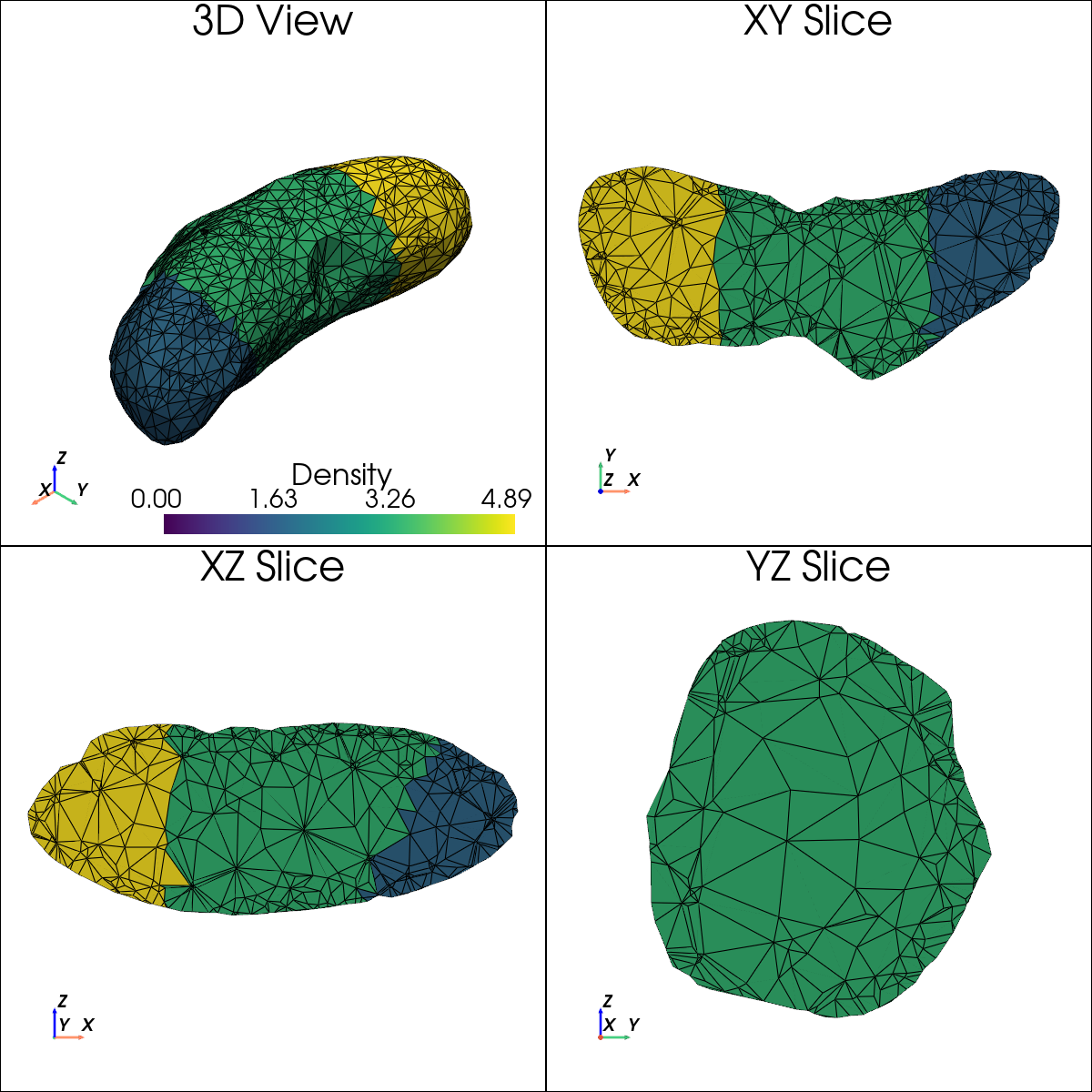}
        \label{fig:mesh_eros_3}
    }
    \subfigure[Eros - Uniform]{
        \includegraphics[width=0.31\linewidth]{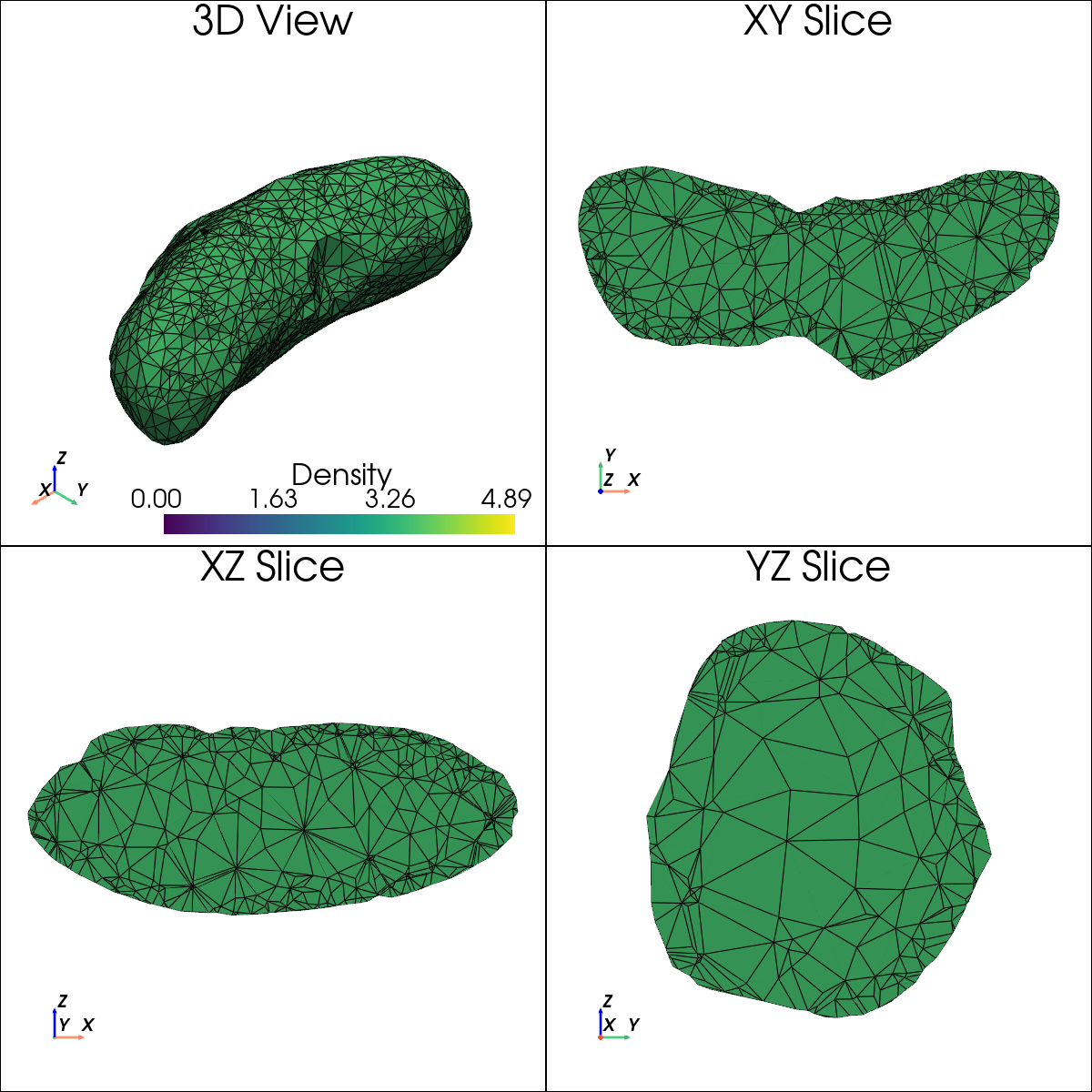}
        \label{fig:mesh_eros_uniform}
    }
    \subfigure[Itokawa - Two regions]{
        \includegraphics[width=0.31\linewidth]{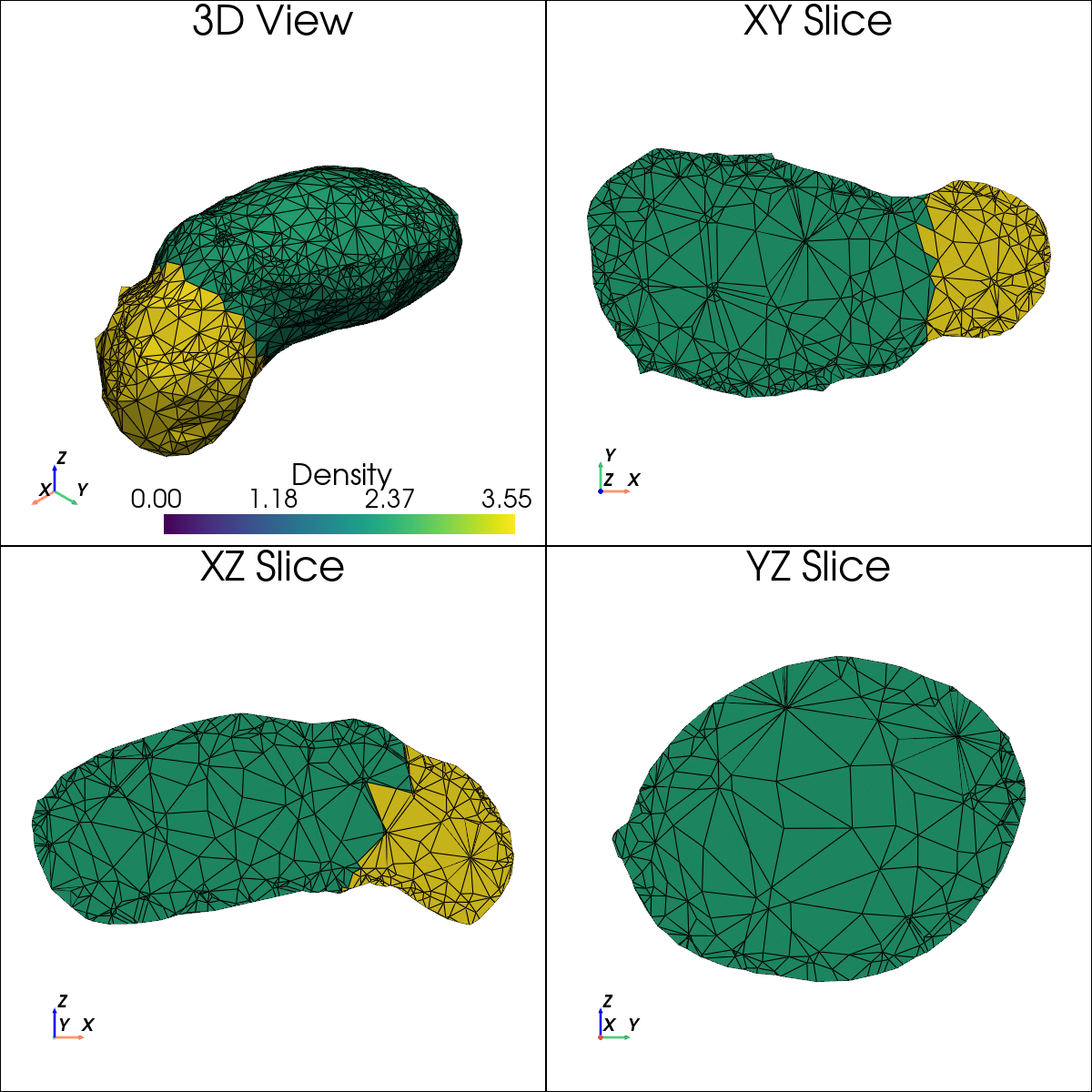}
        \label{fig:mesh_itokawa}
    }
    \subfigure[Itokawa - Smooth]{
        \includegraphics[width=0.31\linewidth]{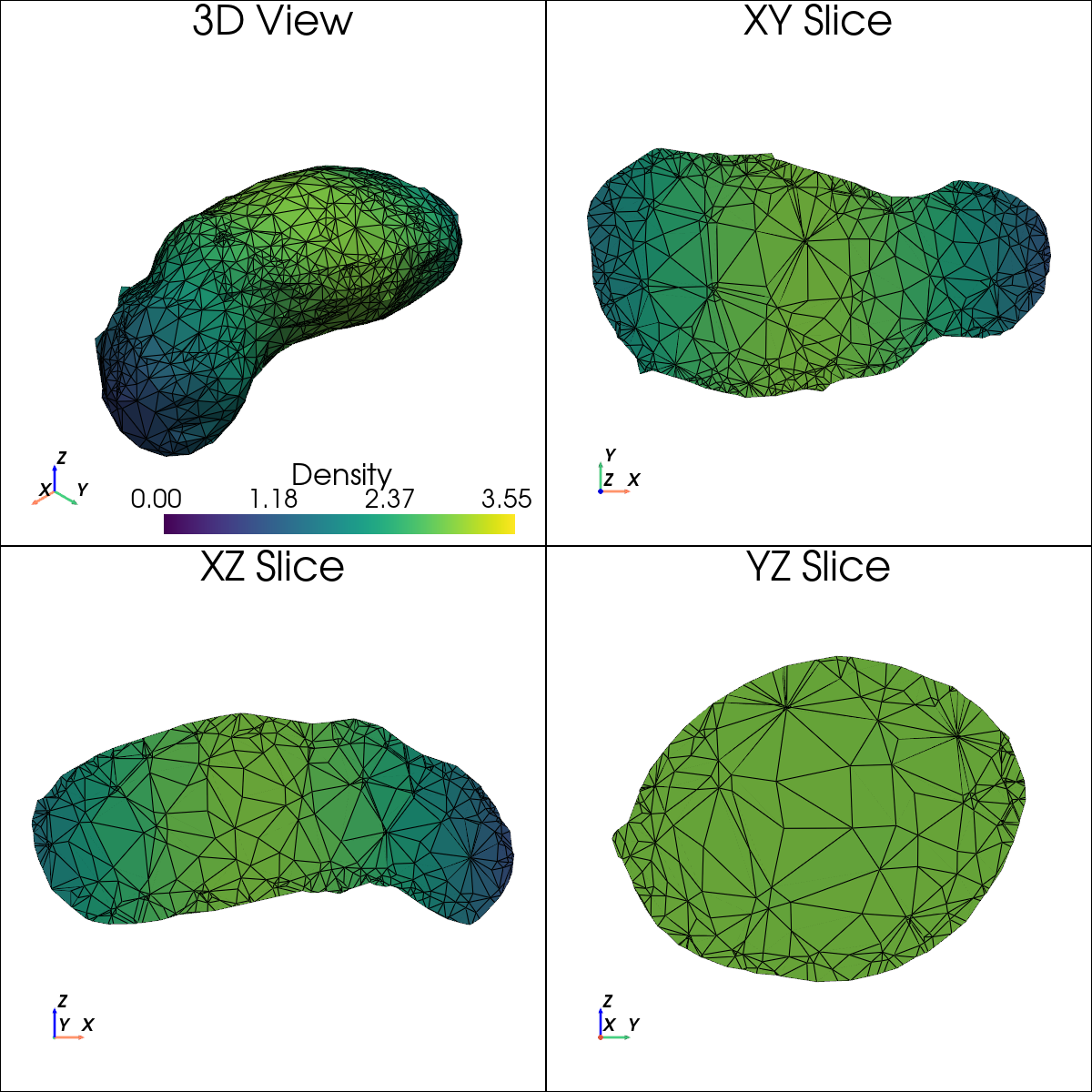}
        \label{fig:mesh_itokawa_smooth}
    }
    \subfigure[Planetesimal - Hollow]{
        \includegraphics[width=0.31\linewidth]{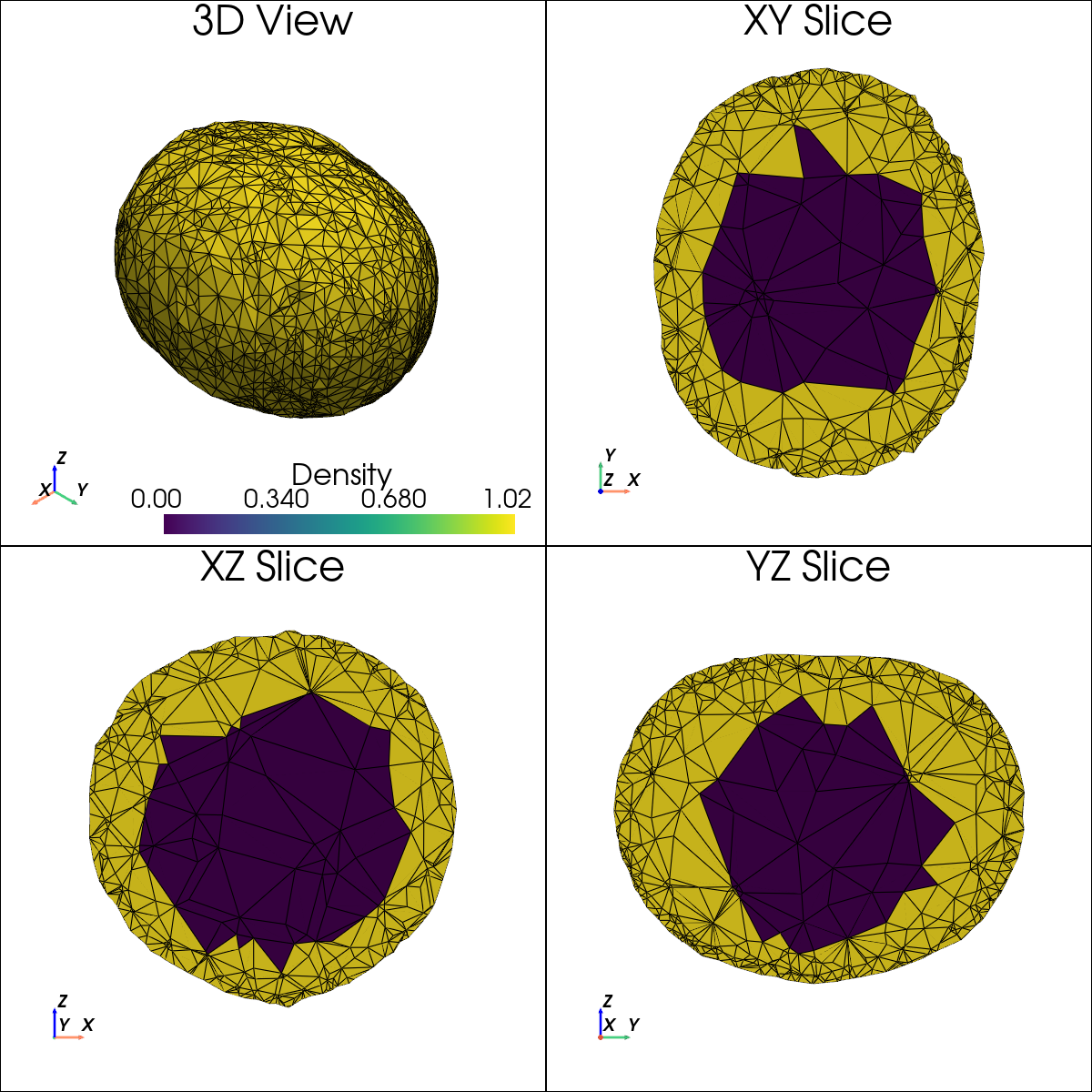}
        \label{fig:mesh_planetesimal}
    }
    \subfigure[Planetesimal - Hollow decentered]{
        \includegraphics[width=0.31\linewidth]{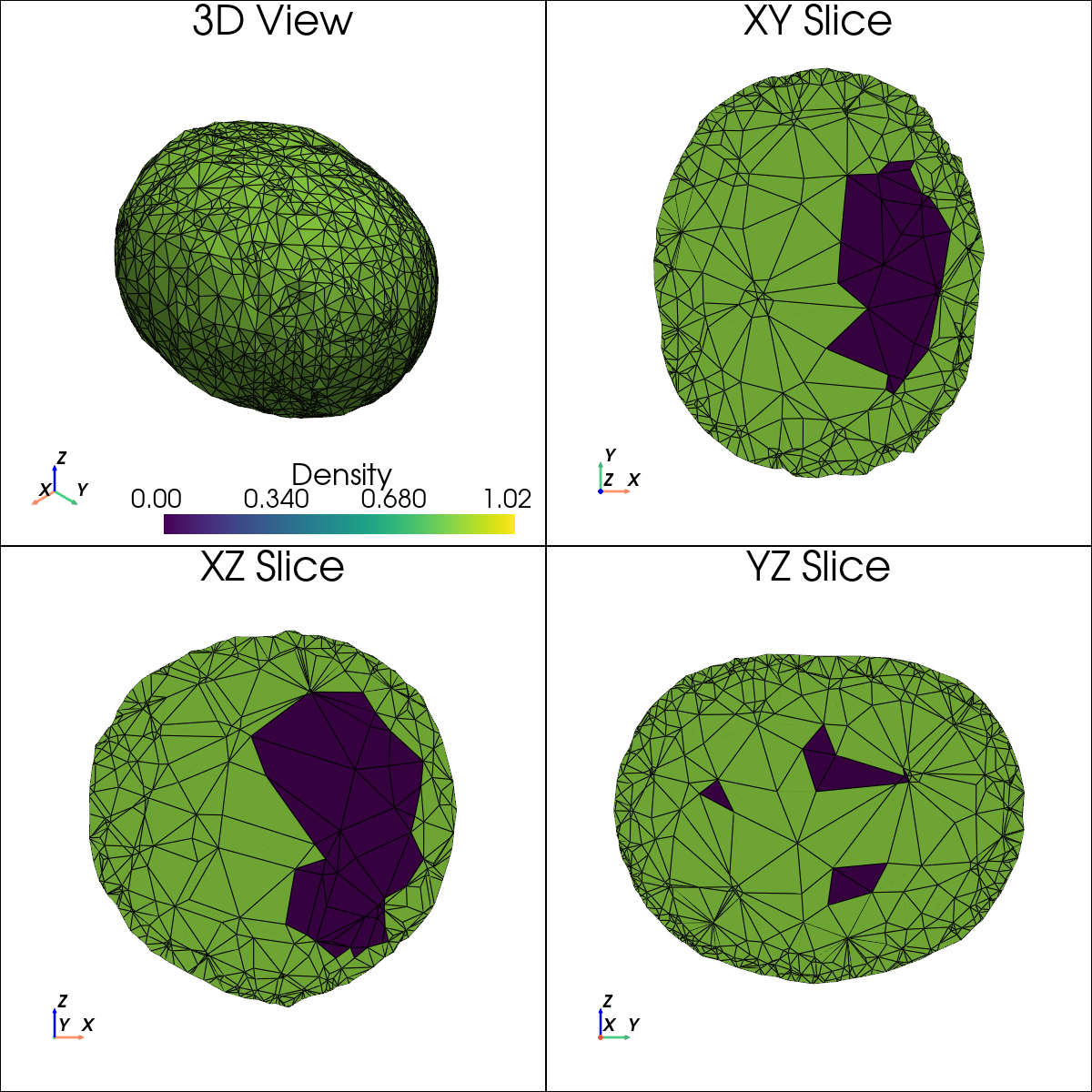}
        \label{fig:mesh_planetesimal_decentered}
    }
    \subfigure[Planetesimal - Uniform]{
        \includegraphics[width=0.31\linewidth]{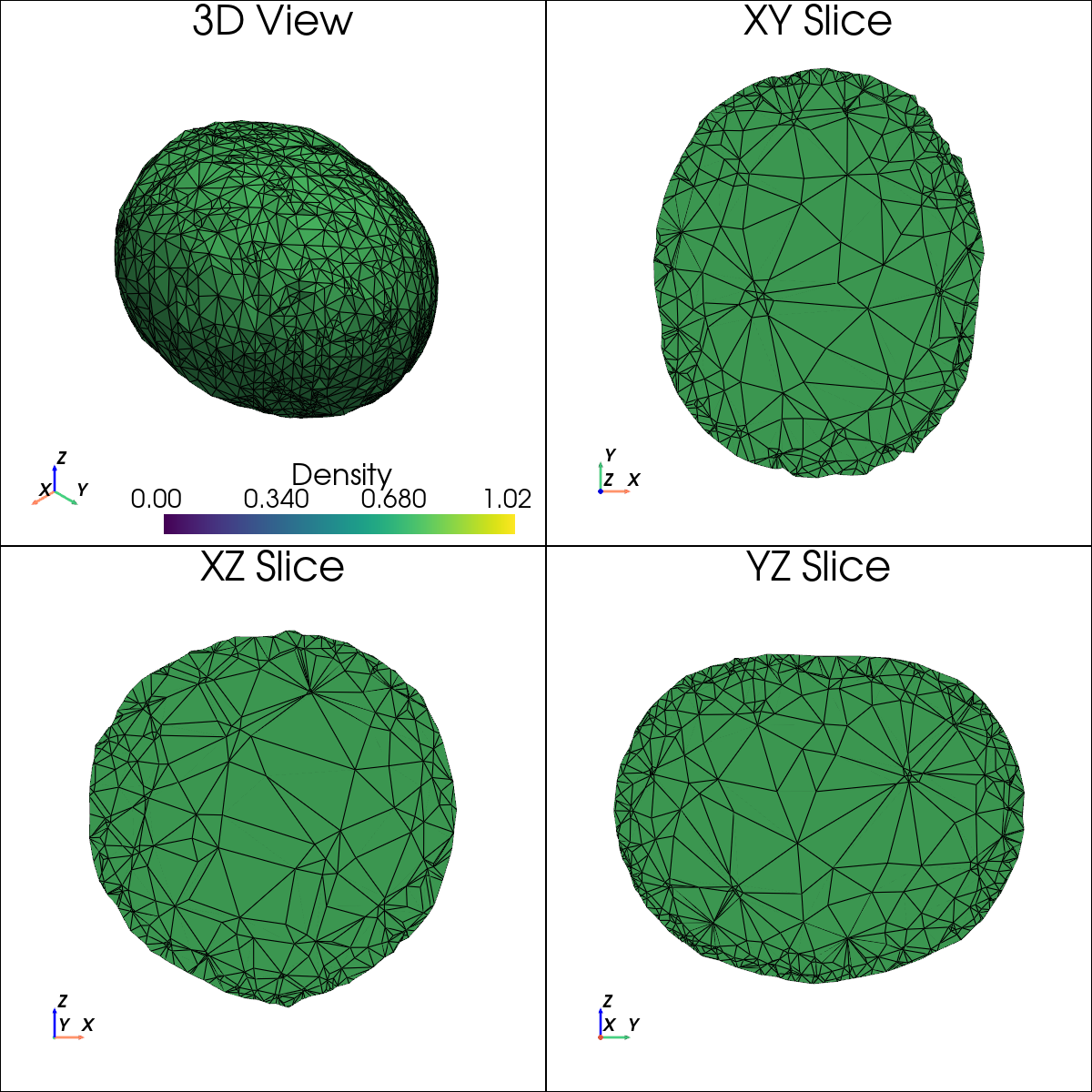}
        \label{fig:mesh_planetesimal_uniform}
    }
    \caption{Density distributions of the ground truth mascon models, with tetrahedron color representing local density. The figure compares several configurations, including regional variations (\subref{fig:mesh_bennu}, \subref{fig:mesh_eros_2}, \subref{fig:mesh_eros_3}, \subref{fig:mesh_itokawa}), hollow structures (\subref{fig:mesh_planetesimal}, \subref{fig:mesh_planetesimal_decentered}), and uniform bodies (\subref{fig:mesh_eros_uniform}, \subref{fig:mesh_planetesimal_uniform}), alongside a model with a smooth density gradient (\subref{fig:mesh_itokawa_smooth}). Each model's total mass is normalized to 1, and models derived from the same base mesh utilize the same color legend to facilitate comparison.}
    \label{fig:meshes}
\end{figure}

To create some fictitious internal heterogeneity, we modified the mass values of selected mascons and subsequently re-normalized the total mass to unity. Bennu was partitioned into three horizontal layers: the outer layers were assigned a higher density, while the central region was given a lower density. For Eros, we studied 3 different versions: a homogeneous one, one partitioned in 2 homogeneous regions, and one partitioned in 3. Itokawa was divided into two regions, following a configuration inspired by its observed geodesy. A second version of Itokawa presents instead a smooth transition between densities, For the fictitious planetesimal, two hollow configurations were created by setting the mass of selected mascons to zero and then normalizing the mascon values so that their sum remained equal to one. In one version, the hollow region is centrally located; in the other, referred to as the \textit{decentered} model, the hollow region is offset from the center. A third version of planetesimal presents a homogeneous distribution. The resulting heterogeneous mascon models are illustrated in Figure~\ref{fig:meshes}, where the tetrahedra are color-coded according to their assigned density.

Details on the discretization process and the mass modifications used to model heterogeneity are provided in the Appendix~\ref{sec:appendix_tetra}.

Our MasconCube model operates on a fixed 3D grid. During training, we only regress the dimensionless mass of each mascon, while its position remains constant. This design choice makes it impossible to perfectly replicate the ground-truth models, as their mass concentrations are not located on a regular grid but at the centroids of the asteroid's tetrahedral mesh elements.

This limitation is intentional. Training our model against a ground truth that also uses a 3D grid would simplify the task artificially and could lead to biased, overly optimistic results. \rev{The key concern is that grid-aligned evaluation would enable perfect spatial correspondence between model predictions and ground truth mass elements, allowing the model to achieve artificially high accuracy rather than developing genuine capability to approximate continuous or irregularly distributed mass fields. By maintaining the physically faithful tetrahedral ground truth derived from real asteroid mesh models, our evaluation ensures that MasconCube must genuinely learn to approximate arbitrarily distributed mass rather than exploiting alignment artifacts. This structural mismatch acts as a stress test: if the model performs well despite the representational incompatibility, it demonstrates robustness that should transfer to operational scenarios where the true internal mass distribution is unknown and certainly not grid-aligned.}

To enable a fair qualitative comparison, we also create a \textit{cubified} version of the ground truth. This process maps the tetrahedral data onto the same 3D grid used by our MasconCube. For each grid point, we first identify the mesh tetrahedron that contains it. We then assign a dimensionless mass to that grid point, calculated specifically so that its implied density matches the density of the enclosing tetrahedron. \rev{Importantly, this cubified representation serves exclusively for visualization purposes and is never used for training or quantitative evaluation.}

\subsection{Training}

\begin{figure}
    \centering
    \includegraphics[width=1.0\linewidth]{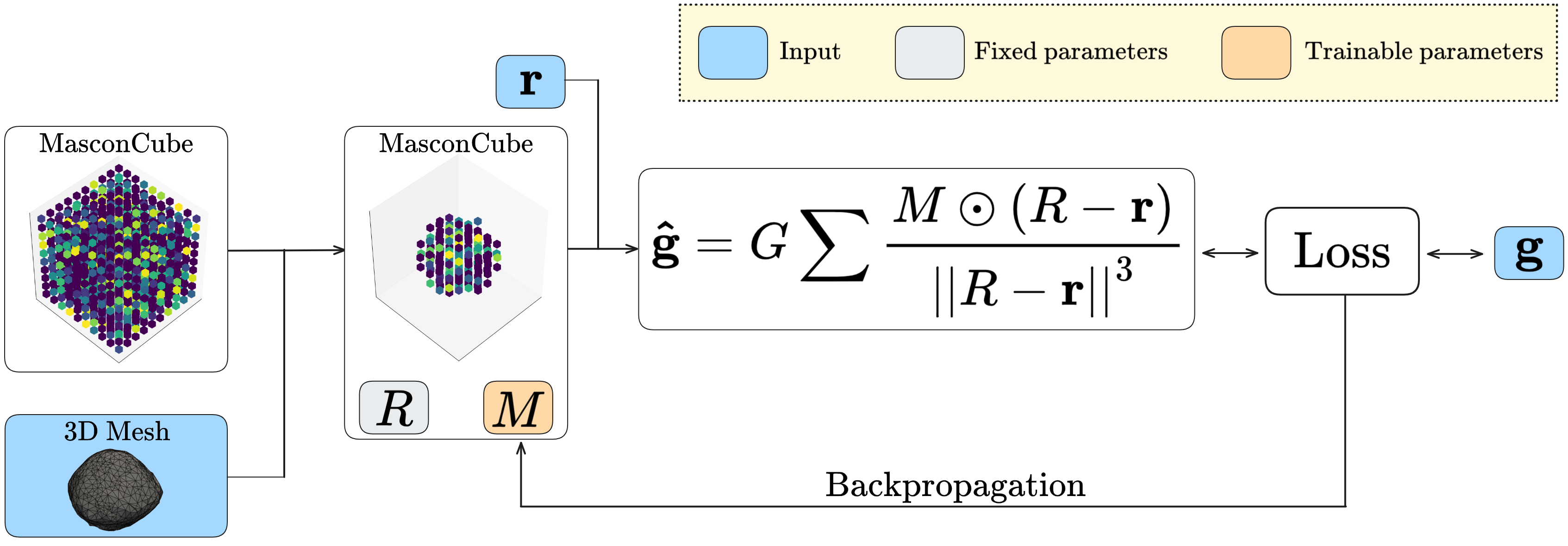}
    \caption{Simplified scheme of the proposed method. A 3D mesh is employed to remove points from a randomly initialized MasconCube. The resulting mascon model predicts an acceleration $\mathbf{\hat{g}}$ at an input position $\mathbf{r}$, which is then compared with the reference value $\mathbf{g}$. The computed loss is backpropagated to update the masses $M$.}
    \label{fig:scheme}
\end{figure}

MasconCubes are trained with a self-supervised learning approach, \rev{a paradigm where supervision signals are derived from the data or physical laws themselves rather than manual annotation}, to reconstruct the internal mass distribution of asteroids from external gravity measurements, as illustrated in Figure~\ref{fig:scheme}.
\rev{It is necessary to train a dedicated model for each asteroid, as the gravity field and internal structure are uniquely determined by the individual body’s shape and mass distribution.}

Unlike traditional geodesy methods, MasconCubes formulates the problem as a direct optimization of mass values in a 3D grid of point masses (mascons). The training code, as well as the code to reproduce all the results in this paper, is provided online~\cite{github}.

\subsubsection{Mascon Grid Initialization}

We define a fixed Cartesian grid of $N^3$ mascons encompassing the target body. Each mascon is characterized by a scalar mass value, initialized uniformly or based on a prior density estimate if available. The spatial resolution of the grid can be adapted depending on the target body's size, desired precision, and computational budget. In our experiments we used MasconCubes of size $100\times100\times100$.

To ensure that mass is only assigned within the physical volume of the asteroid, we utilize the known 3D shape model of the body, typically represented as a surface mesh. After the initial grid is defined, all mascons whose centers fall outside the mesh are removed from the optimization, leaving only $\hat{N}$ mascons, with $\hat{N} < 10^6$. This step constrains the solution space and prevents the model from assigning mass to regions that are known to be empty space, thereby improving both physical realism and computational efficiency. The remaining mascons are initialized with a random mass value sampled from the uniform distribution $U\left(\nicefrac{-1}{10\hat{N}},\nicefrac{1}{10\hat{N}}\right)$.

\rev{While different initialization schemes may affect convergence speed, the current initialization approach was empirically found to generally lead to better convergence behavior compared to alternatives evaluated during preliminary experiments. Furthermore, training with different random seeds consistently results in similar converged models, suggesting that the training process is robust to initialization variability.}

\subsubsection{Observation Model}

\rev{While MasconCubes does not employ a traditional neural network architecture, the model can be understood as a parameter optimization problem where the trainable parameters are the $\hat{N}$ mascon masses.}

The model receives as input a set of gravity observations $\{\mathbf{g}_i\}$ at known spatial positions $\{\mathbf{r}_i\}$, either simulated or derived from spacecraft tracking data. The gravitational acceleration at each observation point is predicted by summing the Newtonian gravity contributions from all mascons:

\begin{equation}
\hat{\mathbf{g}}_i = -G \sum_{j=1}^{N^3} \frac{m_j (\mathbf{r}_i - \mathbf{r}_j)}{\|\mathbf{r}_i - \mathbf{r}_j\|^3}
\end{equation}

where $G$ is the gravitational constant, $m_j$ is the mass of the $j$-th mascon located at position $\mathbf{r}_j$, and $\hat{\mathbf{g}}_i$ is the predicted gravitational acceleration at location $\mathbf{r}_i$.

In our formulation, the mascon masses $\{m_j\}$ are normalized such that their total sum is one, i.e., $\sum_j m_j = 1$. This normalization focuses the learning process on the relative distribution of mass rather than its absolute scale, which is assumed to be known or fixed. \rev{Architecture details are provided in the appendix~\ref{sec:appendix_training}.}

\subsubsection{Loss Function and Optimization}

The model parameters $\{m_j\}$ are optimized to minimize the discrepancy between the predicted and observed gravitational accelerations. Due to the mass normalization constraint, the model cannot freely adjust the magnitude of the predicted acceleration field. To address this, as proposed in previous literature~\cite{izzo2022geodesy}, we use a normalized L1 loss that compensates for global scale differences by introducing an optimal scalar factor $c$:

\begin{equation}
\label{eq:loss}
\mathcal{L} = \frac{1}{M} \sum_{i=1}^{M} \| \mathbf{g}_i - c \hat{\mathbf{g}}_i \|_1, \quad \text{where} \quad c = \frac{\sum_i \mathbf{g}_i \cdot \hat{\mathbf{g}}_i}{\sum_i \|\hat{\mathbf{g}}_i\|^2}
\end{equation}

where $\hat{\mathbf{g}}_i$ is the predicted acceleration from the normalized mass distribution, and $c$ is a scalar scaling factor that adjusts the global magnitude of the field.

The model is trained for 1000 steps, taking in input each time a batch of 1000 positions randomly sampled inside a sphere of radius 1 paired with their ground-truth accelerations. The batch is changed every 10 epochs and Adam~\cite{kingma2017adam} is used to update the trainable mass values, with a starting learning rate equals to \num{1e-5} reduced using step decay. Training parameters are provided in detail in the appendix~\ref{sec:appendix_training}.

\subsection{Evaluation}

To ensure a comprehensive evaluation of the learned mass distribution and provide a fair comparison with existing literature, we employ multiple complementary assessment methodologies that highlight the strengths and weaknesses of different approaches across various use cases. Our evaluation framework encompasses both qualitative and quantitative analyses, including visual comparison of internal density distributions, spherical harmonic coefficient analysis through normalized Stokes coefficients, gravitational acceleration prediction accuracy at multiple altitudes, and trajectory simulation performance for realistic spacecraft dynamics. Results and comparisons are provided in Section~\ref{sec:results}, while this section serves only to provide the theoretical background of the metrics used.

\subsubsection{Spherical Harmonics Analysis}

As gravitational and geodetic representation are strongly related, we compare the normalized Stokes coefficients of the spherical harmonic expansion of the learned model with the ground truth. We use the formal definition for the description of the gravitational potential $U$ from \cite{tricarico2013global}:

\begin{equation}
    U(r,\theta,\phi) = \frac{GM}{r} \sum_{l=0}^{l=\infty} \sum_{m=0}^{m=l} \left( \frac{R_0}{r} \right)^l P_{lm} \left( \cos\theta \right) \cdot \left( C_{lm} \cos m\phi + S_{lm} \sin m\phi \right)
\end{equation}
where $G$ is the universal gravitational constant, $M$ is the total mass of the body, $r$, $\theta$, and $\phi$ are the spherical coordinates in the body-fixed reference frame, $R_0$ is the reference radius, $P_{lm}(x)$ is the zonal Legendre polynomial, and $C_{lm}$ and $S_{lm}$ are respectively the cosine and sine Stokes coefficients, that can be determined from the density distribution $\rho(r,\theta,\phi)$ integrating over the volume $V$ of the body:
\begin{equation}
    \begin{Bmatrix}C_{lm}\\S_{lm}\end{Bmatrix} = \frac{(2-\delta_{m,0})}{M} \frac{(l-m)!}{(l+m)!} \int_V \rho(r,\theta,\phi) \left(\frac{r}{R_0}\right)^l \cdot P_{lm}(\cos\theta) \begin{Bmatrix} \cos m\phi \\ \sin m\phi\end{Bmatrix} \,dV
\end{equation}
which in a mascon model defined by the set $\mathcal{M} = \{(m_i,r_i,\phi_i,\theta_i)\}_{i=1}^N$ can be approximated with

\begin{equation}
    \begin{Bmatrix}C_{lm}\\S_{lm}\end{Bmatrix} \approx \frac{(2-\delta_{m,0})}{M} \frac{(l-m)!}{(l+m)!} \sum_{i=1}^N m_i \left(\frac{r_i}{R_0}\right)^l \cdot P_{lm}(\cos\theta_i) \begin{Bmatrix} \cos m\phi_i \\ \sin m\phi_i\end{Bmatrix}
\end{equation}

$C_{lm}$ and $S_{lm}$ are then normalized by the normalization factor $N_{lm}$ defined as:
\begin{equation}
    N_{lm} = \sqrt{\frac{(l+m)!}{(2-\delta_{m,0})(2l+l)(l-m)!}}
\end{equation}
so that our normalized Stokes coefficients are $\{\tilde{C}_{lm}, \tilde{S}_{lm}\} = N_{lm}\left\{C_{lm}, S_{lm}\right\}$.

For this study we use Stokes coefficients up to the seventh degree and we do not consider $\tilde{S}_{lm}$ when $l=m$ because they are always equal to zero.

To compare how different methods approximate the gravity field both at a general and a degree-level, we compute the Mean Absolute Error (MAE) up to the $D$th degree defined as:
\begin{equation}
    \label{eq:mae}
    \text{MAE}_D = \frac{1}{(D+1)^2}\sum_{l=0}^D \sum_{m=0}^l \left( |\Delta \tilde{C}_{lm}| + |\Delta \tilde{S}_{lm}|  \right)
\end{equation}
where $\Delta \tilde{C}_{lm} = \tilde{C}_{lm}^{\text{ predicted}} - \tilde{C}_{lm}^{\text{ truth}}$ and $\Delta \tilde{S}_{lm} = \tilde{S}_{lm}^{\text{ predicted}} - \tilde{S}_{lm}^{\text{ truth}}$.

\section{Results}
\label{sec:results}

In this section, we compare MasconCubes with GeodesyNets, PINN-GM-III, and the \textit{cubified} ground-truth, created as described in Section~\ref{sec:asteroids}, which serves as our gold standard. PINN-GM-III have been re-implemented following the details provided in the original paper to evaluate them on our ground-truth models. For completeness, we also report the original metrics from the paper, in case our re-implementation deviates from the intended performance. To ensure a comprehensive and fair comparison that highlights the strengths and weaknesses of each method, we evaluate them using several metrics. 

\rev{MasconCubes require the asteroid's shape model to be known \textit{a priori}. To ensure fair comparison, we exclusively compare against the differential training version of GeodesyNets, which also exploits shape information to improve accuracy and efficiency. While PINN-GM III cannot exploit \textit{a priori} knowledge of the shape of studied body, it requires the Stokes coefficient $\tilde{C}_{20}$ as prior knowledge. Thus, all three methods leverage geometric priors. Shape requirement is practically realistic: in asteroid exploration missions, shape models are typically obtained from optical imagery or radar during approach phases, before high-fidelity gravity data becomes available from close-proximity tracking.}

\subsection{Qualitative Comparison of the Geodetic Models}

To qualitatively compare the learned geodetic models, we visualize their resulting mass or density distributions. For MasconCubes and the ground-truth Mascon Model, we can directly visualize their mass distributions as they are explicit representations, however being the ground-truth models sparse, their visualization is not particularly insightful, for this reason we visualize their \emph{cubified} version. On the other hand, GeodesyNets provide an implicit density representation, so we visualize their learned distribution by evaluating the model at numerous points inside the body.

PINN-GM-III models do not directly represent the body's geodesy; instead, they learn an implicit representation of the gravitational potential. However, the density profile can be reconstructed by evaluating Poisson's equation inside the body. We use this method, a possibility suggested but not explored in the original paper, to visualize the density distribution learned by PINN-GM-III.

\begin{figure}
    \centering
    \subfigure[\emph{Cubified} ground-truth]{
        \includegraphics[width=0.48\linewidth]{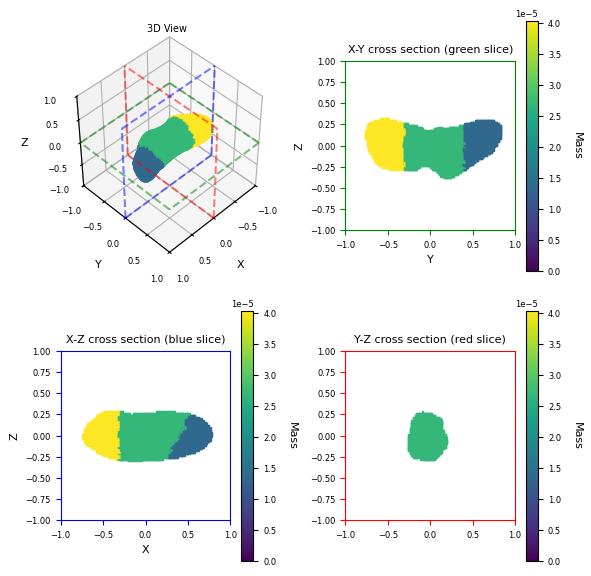}
        \label{fig:density_eros3_gt}
    }
    \hfill
    \subfigure[MasconCube]{
        \includegraphics[width=0.48\linewidth]{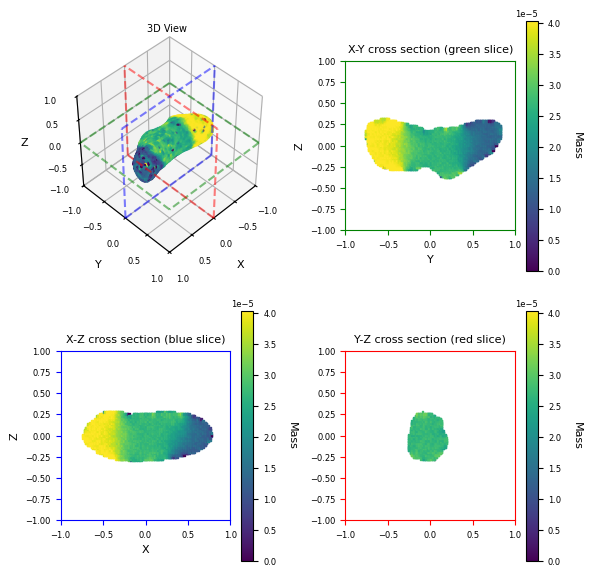}
        \label{fig:density_eros3_cube}
    }
    \subfigure[GeodesyNet]{
        \includegraphics[width=0.48\linewidth]{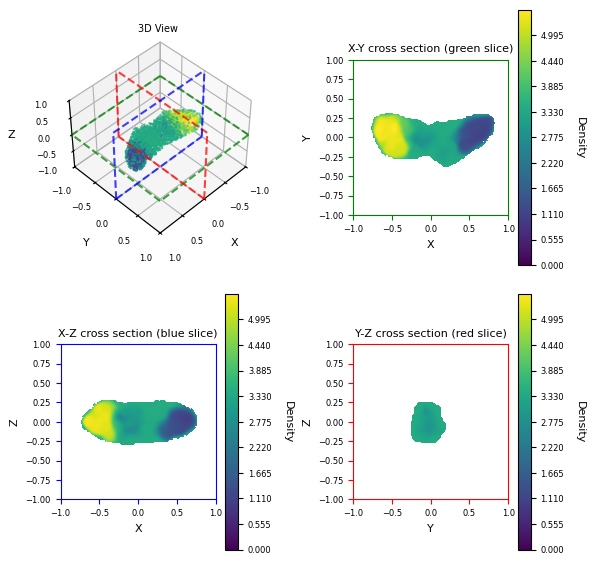}
        \label{fig:density_eros3_net}
    }
    \hfill
    \subfigure[PINN-GM~III]{
        \includegraphics[width=0.48\linewidth]{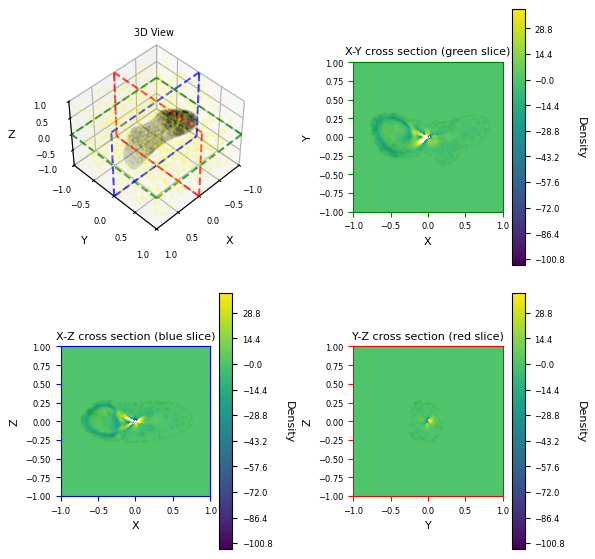}
        \label{fig:density_eros3_pinn}
    }
    \caption{Internal structure comparison for \emph{Eros – Three Regions} model across different methods.}
    \label{fig:density_eros3}
\end{figure}

Figure~\ref{fig:density_eros3} shows the learned distribution for \emph{Eros - Three regions} and a plot of the cubified ground truth for comparison. While both the GeodesyNet and the MasconCube learn an internal structure similar to the target one, the density distribution learned by PINN-GM III is completely different, with a concentration of density around the center of mass of the center of mass of the asteroid, and almost zero density everywhere else. This behavior of PINN-GM have been observed for all the asteroids in the dataset, leading us to conclude that, while it remains a valid approach to estimate the gravitational acceleration, it is not suited to study the geodesy of a body.

\begin{figure}
    \centering
    \subfigure[\emph{Cubified} ground-truth]{
        \includegraphics[width=0.31\linewidth]{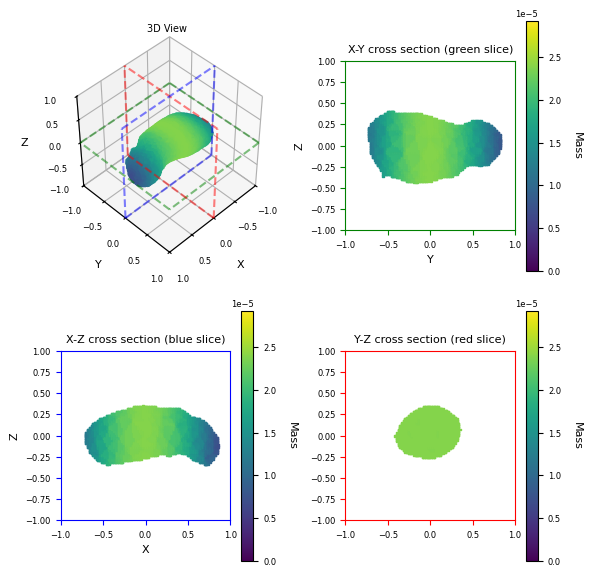}
        \label{fig:density_itokawacos_gt}
    }
    \hfill
    \subfigure[MasconCube]{
        \includegraphics[width=0.31\linewidth]{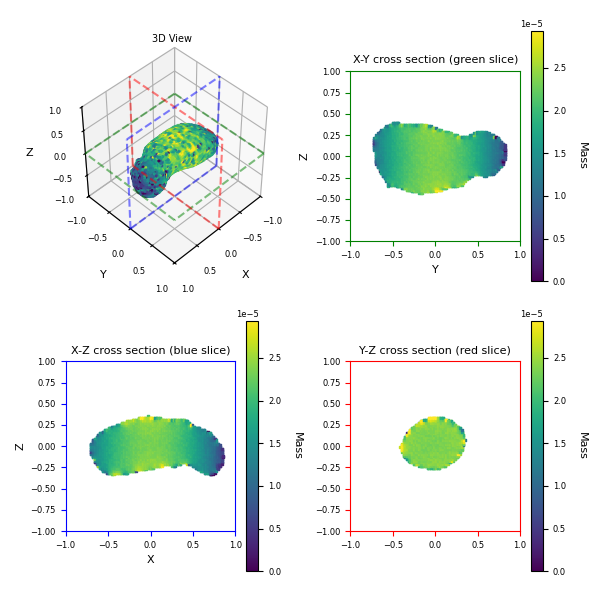}
        \label{fig:density_itokawacos_cube}
    }
    \hfill
    \subfigure[GeodesyNet]{
        \includegraphics[width=0.31\linewidth]{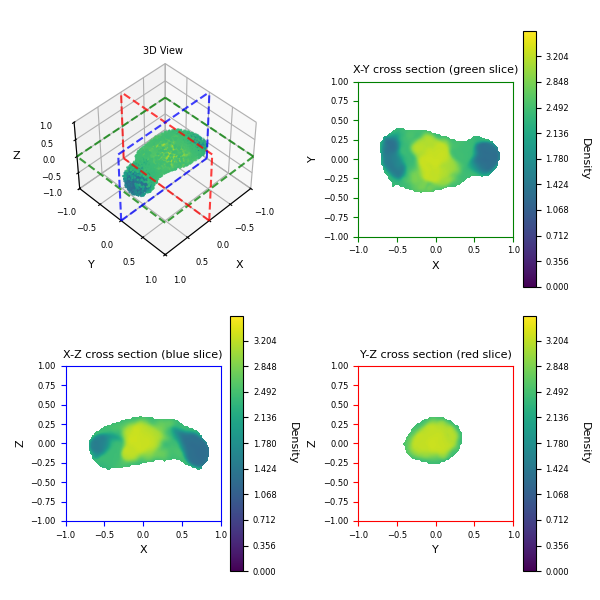}
        \label{fig:density_itokawacos_net}
    }
    \caption{Internal structure comparison for \emph{Itokawa – Smooth} model across different methods.}
    \label{fig:density_itokawacos}
\end{figure}

An important difference between the methods is observed in their ability to model smooth density distributions. Figure~\ref{fig:density_itokawacos} compares the internal structure of \textit{Itokawa - Smooth} as learned by MasconCube and GeodesyNet, demonstrating that MasconCube produces a more accurate representation of the smooth density variations.

\subsection{Stokes Coefficients}

\begin{figure}
    \centering
    \subfigure[MasconCube Stokes Coefficients errors]{
        \includegraphics[width=0.47\linewidth]{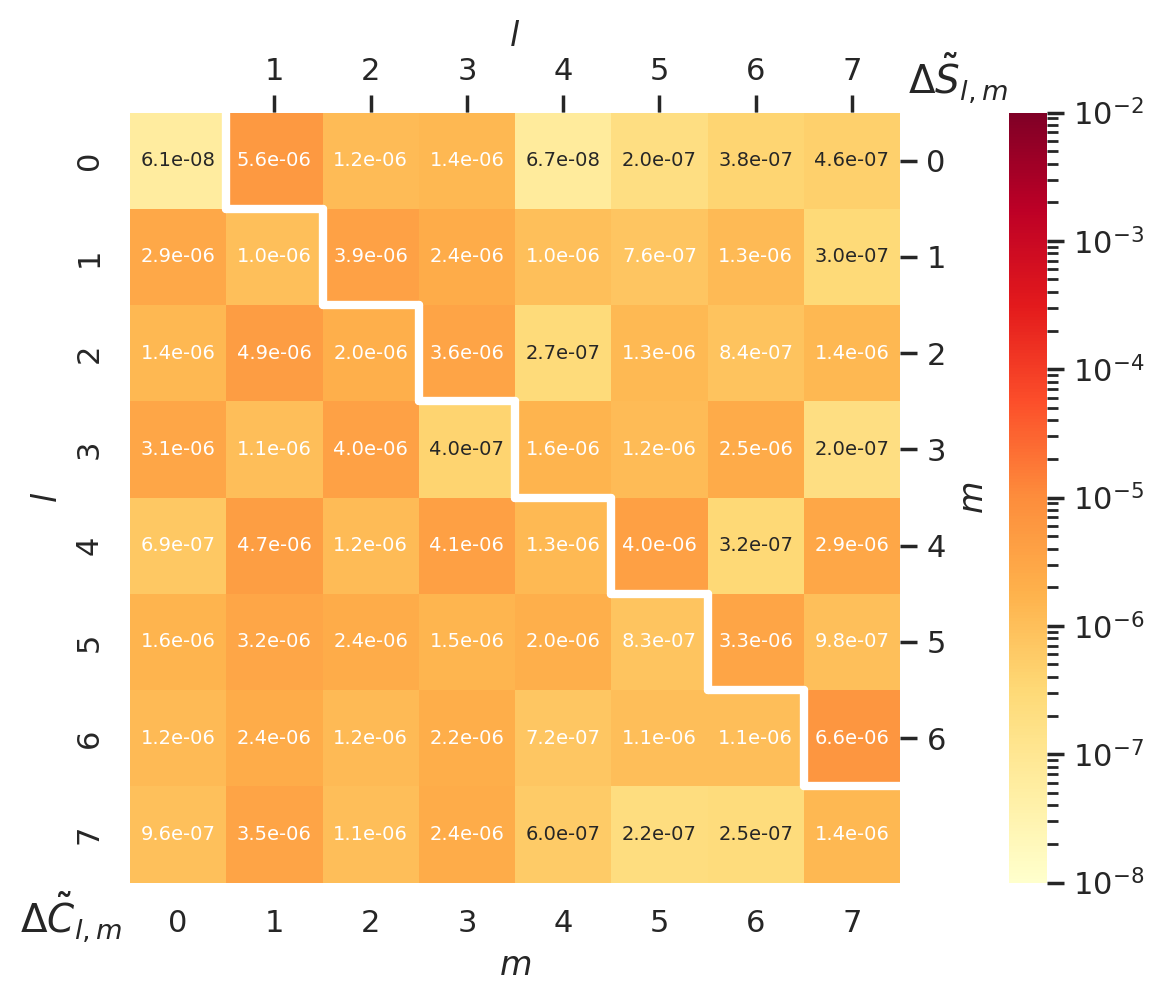}
        \label{fig:stokes_cube}
    }
    \hfill
    \subfigure[GeodesyNet Stokes Coefficients errors]{
        \includegraphics[width=0.47\linewidth]{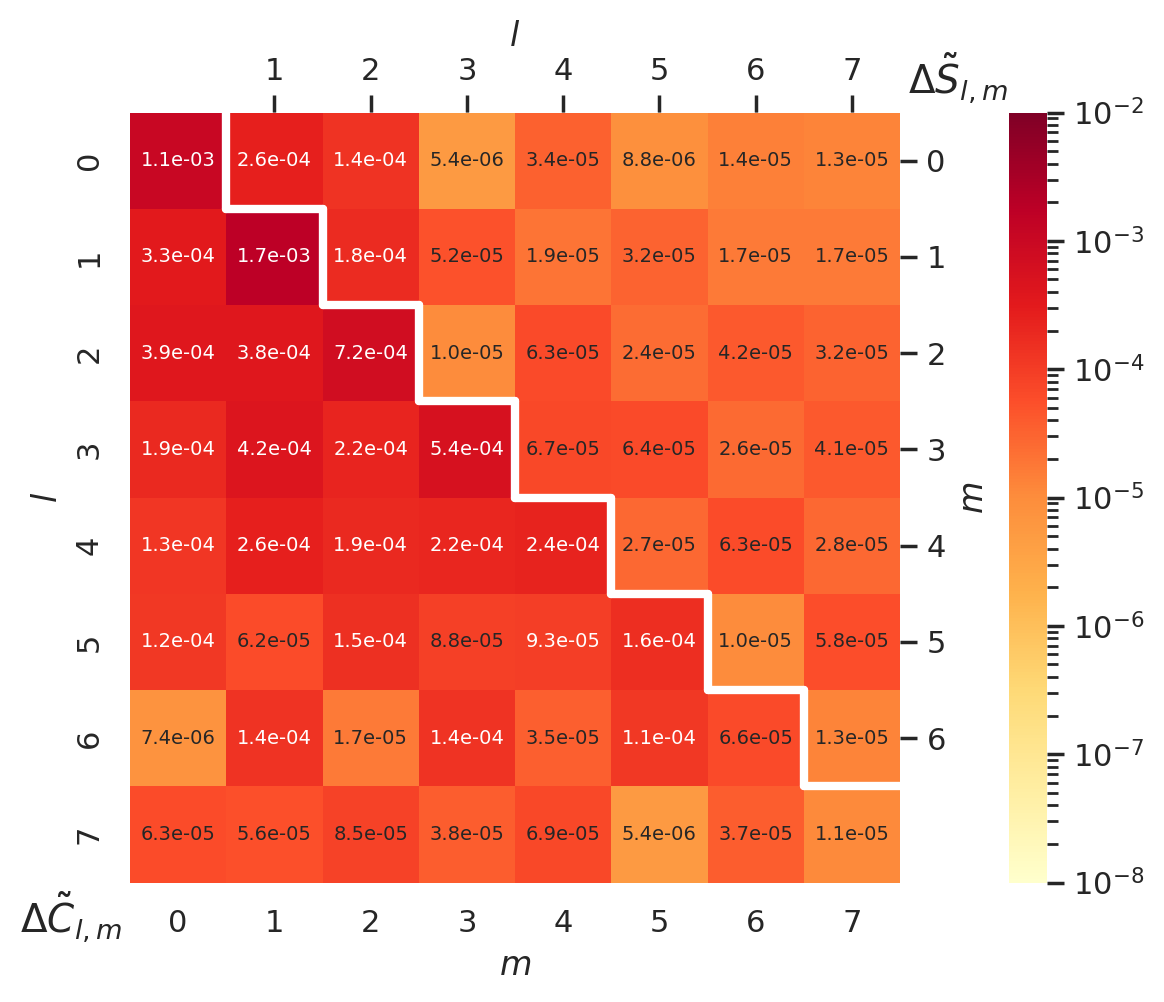}
        \label{fig:stokes_net}
    }
    \subfigure[Mean Absolute Error \(\text{MAE}_n\) of Stokes Coefficients by Degree]{
        \includegraphics[width=0.95\linewidth]{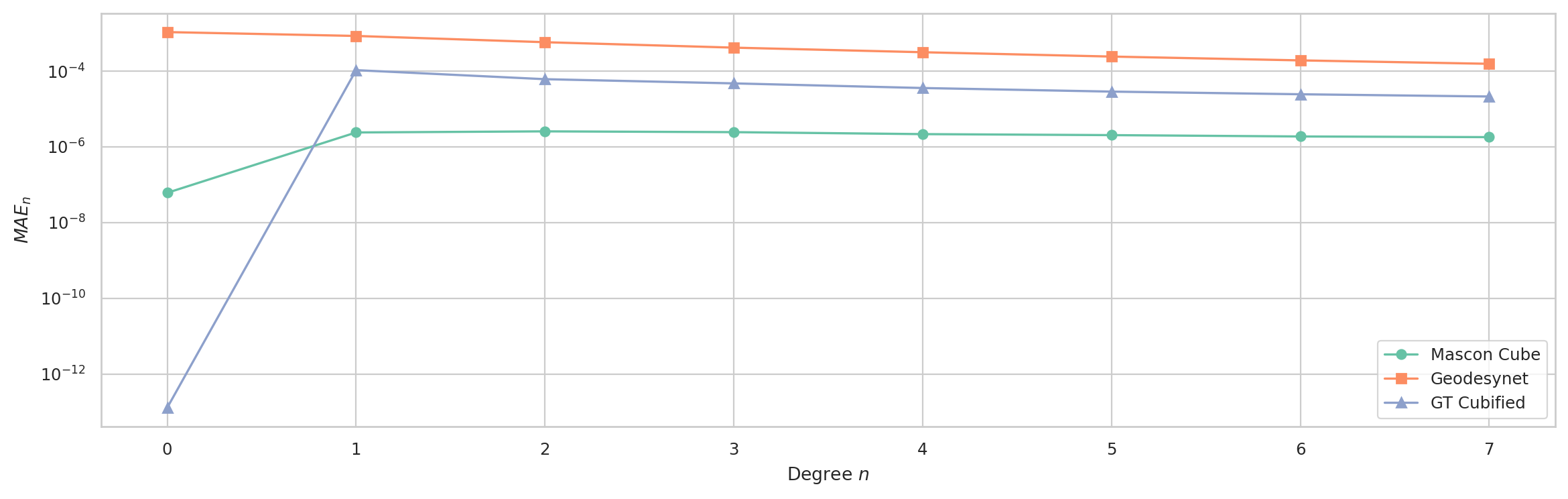}
        \label{fig:stokes_mae}
    }
    \caption{Comparison of normalized Stokes coefficients for the \emph{Eros – Two Regions} model. (a) and (b) show visualizations of the Stokes coefficients obtained from MasconCube and GeodesyNet respectively, illustrating the closer alignment of MasconCube coefficients with the ground truth. (c) plots the Mean Absolute Error \(\text{MAE}_n\) of the coefficients as a function of degree \(n\). The plot highlights MasconCube's consistently lower error across degrees, confirming its superior fidelity in modeling the gravitational potential.}
    \label{fig:stokes}
\end{figure}

Table~\ref{tab:mae} compares the $\text{MAE}_7$ of the normalized Stokes coefficients, as defined in Equation~\ref{eq:mae}, predicted by MasconCube, GeodesyNet, and Cubified ground-truth. A more in detail coefficient-wise comparison for \textit{Eros - Two Regions} is reported in Figure~\ref{fig:stokes}. Overall, the Stokes coefficients analysis reinforces the effectiveness of MasconCube in accurately modeling complex gravitational fields. The smaller MAE values and closer alignment with ground-truth coefficients underscore MasconCube's capability to capture subtle features of the gravitational potential up to high spherical harmonic degrees. This enhanced precision is critical for applications requiring detailed gravitational modeling, such as spacecraft navigation and geophysical studies of irregularly shaped celestial bodies.

\begin{table}
    \centering
    \caption{\rev{Mean Absolute Error (MAE) of normalized Stokes coefficients up to degree 7 for different gravitational models and bodies. PINN-GM-III is excluded due to its divergent internal structure (see Figure~\ref{fig:density_eros3}). Lowest values per body are in bold.}}
    \begin{tabular}{
        l
        S[table-format=1.3e-1]
        S[table-format=1.3e-1]
        S[table-format=1.3e-1]
    }
        \toprule
        \multirow{2}{*}{\textbf{Body}} & \multicolumn{3}{c}{\textbf{MAE\(_7\)}} \\
        & {\medspace\medspace\textbf{MasconCube}\medspace\medspace} & {\medspace\textbf{GeodesyNet}\medspace} & {\textbf{Cubified Ground Truth}} \\
        \midrule
        Bennu & \num{1.899e-04} & \num{2.292e-04} & \textbf{\num{2.114e-05}} \\ 
        Eros - Two regions & \textbf{\num{2.364e-06}} & \num{1.081e-04} & \num{2.452e-05} \\
        Eros - Three regions & \textbf{\num{7.734e-06}} & \num{1.704e-04} & \num{2.041e-05} \\
        Eros - Uniform & \textbf{\num{2.037e-06}} & \num{5.348e-05} & \num{2.661e-05} \\
        Itokawa - Two regions & \textbf{\num{1.793e-06}} & \num{1.548e-04} & \num{2.121e-05} \\
        Itokawa - Smooth & \num{5.007e-06} & \num{1.405e-04} & \textbf{\num{1.378e-05}} \\
        Planetesimal - Hollow & \textbf{\num{1.594e-06}} & \num{7.956e-05} & \num{1.962e-05} \\
        Planetesimal - Hollow decentered & \textbf{\num{1.399e-05}} & \num{7.853e-05} & \num{1.825e-05} \\
        Planetesimal - Uniform & \textbf{\num{1.408e-06}} & \num{4.100e-05} & \num{1.911e-05} \\
        \bottomrule
    \end{tabular}
    \label{tab:mae}
\end{table}

\subsection{Acceleration}

A reliable gravity model must accurately estimate gravitational acceleration around the target body. For a mascon model $\mathcal{M} = \{(m_i, \mathbf{r}_i)\}_{i=1}^N$, the acceleration at position $\mathbf{p}$ is given by the superposition of point-mass contributions:
\begin{equation}
    \mathbf{a}(\mathbf{p}) = -G\sum_{i=0}^N m_i \frac{\mathbf{p}-\mathbf{r_i}}{||\mathbf{p}-\mathbf{r_i}||^3}
\end{equation}
where $G$ is the gravitational constant.

Table~\ref{tab:acc} reproduces results reported in previous works, mainly from~\cite{martin2025pinn}, to provide a broader context. Since \rev{the reported acceleration errors} come from evaluations on different datasets and under different experimental setups, they are not directly comparable.  
The MasconCube results, along with GeodesyNet trained using the differential framework, were obtained in this work following our own training and evaluation protocol on the \emph{Eros – Two Regions} model. This choice yields a dataset conceptually similar to that of PINN-GM~III, as both are Eros-inspired. Nonetheless, the two differ, since their ground‑truth gravity fields are synthetic and were generated using different procedures. Additionally, we trained a reduced-complexity MasconCube variant, called MasconCube-S \rev{(gride size equal to 40)}, using a similar amount of parameters and training data as PINN-GM~III, to facilitate a direct comparison.

\begin{table}
    \centering
    \caption{\rev{Comparison of machine learning-based gravity models. Most values are taken from the literature (mainly~\cite{martin2025pinn}), while underlined models are computed in this work.}}
    \begin{tabular}{lS[table-format=8.0]S[table-format=8.0]S[table-format=1.3]cc}
        \toprule
         {\textbf{Model}} & {\textbf{Parameters}} & {\textbf{Training Data}} & {\textbf{Avg. Error (\%)}} & {\textbf{Valid Globally}} & {\textbf{Requirements}}  \\
         \midrule
         GP \cite{gao2019gp} & 12960000 & 3600 & 1.5 & \ding{55} & --- \\
         NNs \cite{cheng2019nn} & 1300000 & 800000 & 0.35 & \ding{55} & --- \\
         ELMs \cite{furfaro2021elm} & 350000 & 768000 & 1 & \ding{55} & --- \\
         GeodesyNet \cite{izzo2022geodesy} & 80800 & 1000000 & 0.36 & \ding{51} & --- \\
         \underline{Diff. GeodesyNet} \cite{izzo2022geodesy} & 80800 & 1000000 & 0.116 & \ding{51} & Shape \\
         PINN-GM III \cite{martin2025pinn} & 2211 & 4096 & 0.30 & \ding{51} & $\tilde{C}_{20}$ \\
         \underline{MasconCube} & 35603 & 100000 & 0.006 & \ding{51} & Shape \\
        \underline{MasconCube-S} & 2160 & 4000 & 0.028 & \ding{51} & Shape \\
         \bottomrule
    \end{tabular}
    \label{tab:acc}
\end{table}

A comprehensive and equitable cross-model evaluation is detailed in Table~\ref{tab:acc_geodesynet}, where GeodesyNet was re-trained using its differential training framework to ensure consistent datasets and training conditions across all models. This setup permits a direct and fair comparison with MasconCube. For reference, we include metrics for a uniform density model as a baseline, and a \emph{cubified} ground-truth model, serving as the gold standard for accuracy.

\rev{The comparison with the \emph{cubified} ground-truth makes emerge an apparent discrepancy between superior Stokes coefficient fidelity in Table~\ref{tab:mae} and higher acceleration errors in Table~\ref{tab:acc_geodesynet}. This actually stems from fundamental differences in how these metrics evaluate field fidelity. Stokes coefficients measure the harmonic decomposition of the gravitational potential globally. Conversely, acceleration prediction directly evaluates point-wise field values.}

\begin{figure}
    \centering
    \includegraphics[width=1.0\linewidth]{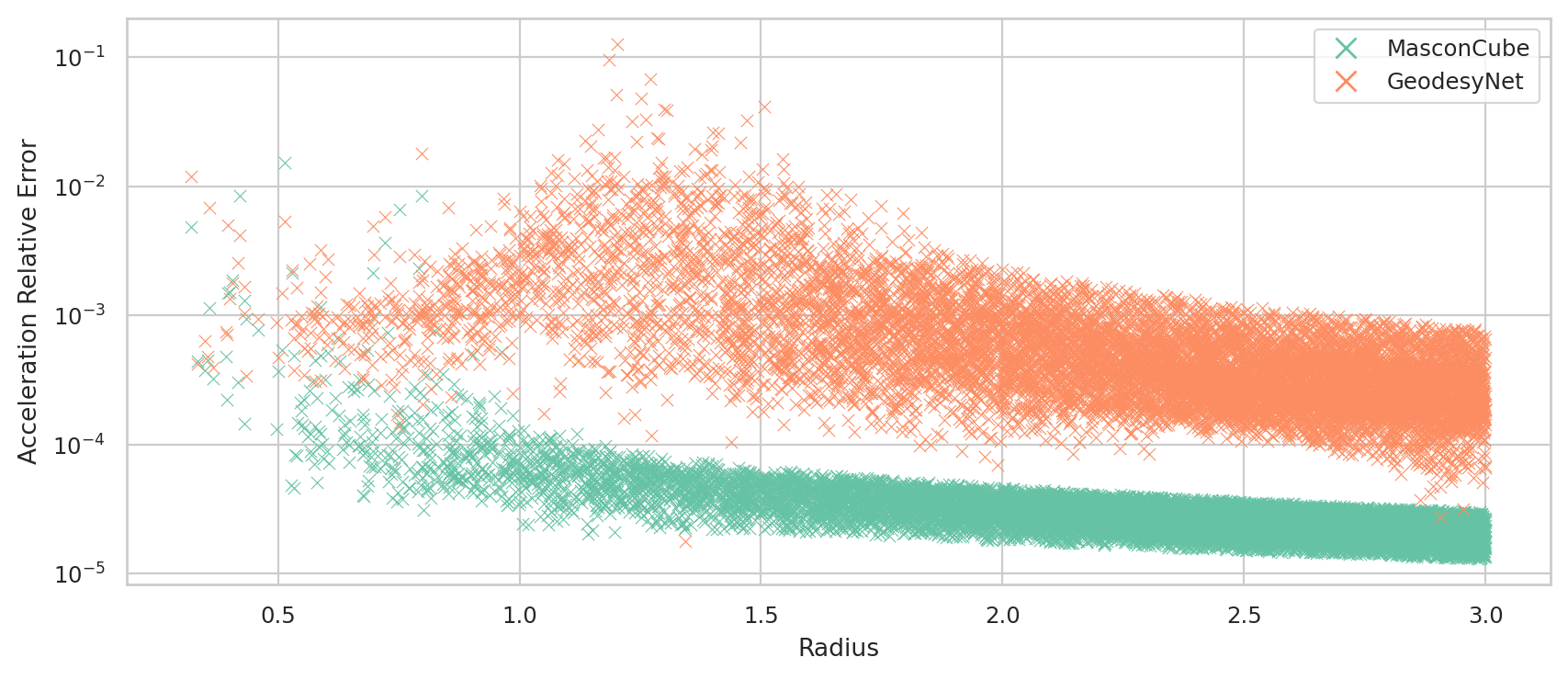}
    \caption{Relative Euclidean Norm Distance of gravitational acceleration estimates at various distances from the coordinate origin used in the \emph{Eros – Two Regions} dataset, comparing MasconCube and GeodesyNet models. The plot shows that MasconCube consistently yields lower relative errors, underscoring its superior accuracy across the analyzed distance range.}
    \label{fig:acc_eros}
\end{figure}

Figure~\ref{fig:acc_eros} illustrates the Relative Euclidean Norm Distance of the gravitational acceleration predictions at various altitudes above the surface of Eros for both GeodesyNet and MasconCube models trained on the \emph{Eros – Two Regions} dataset. The figure highlights MasconCube's superior accuracy and stability across different altitudes, demonstrating its effectiveness in capturing the gravitational field of irregular bodies.

\begin{table}
\centering
\caption{\rev{Evaluation metrics for gravitational acceleration prediction on the \emph{Eros – Two Regions} dataset.}}
\begin{tabular}{lS[table-format=1.2e-1]S[table-format=1.2e-1]S[table-format=1.2e-1]}
\hline
\textbf{Model} & {\textbf{Cosine Distance}} & {\textbf{Euclidean Norm Distance}} & {\textbf{Relative Euclidean Norm Distance}} \\
\hline
MasconCube             & 1.41e-6 & 4.14e-4 & 4.43e-4 \\
GeodesyNet             & 4.02e-5 & 1.97e-3 & 3.98e-3 \\
Uniform model          & 5.42e-4 & 1.92e-2 & 2.66e-2 \\
Cubified ground truth  & 4.60e-7 & 2.72e-4 & 2.56e-4 \\
\hline
\end{tabular}
\label{tab:acc_geodesynet}
\end{table}

\subsection{Trajectory Simulation}

A key application of gravity models is to predict the future trajectory of an object orbiting a celestial body, which is essential for designing fly-by and exploration missions. For each asteroid in our database, we simulated the trajectory of a spacecraft with a mass of 12 kg and a cross-sectional area of 1 m\textsuperscript{2}. The spacecraft’s dynamics were influenced by the gravitational field of the asteroid as well as solar radiation pressure.

Asteroid parameters such as mass, rotation period, aphelion and perihelion distances, and the longest dimension were obtained from real observational data. For theoretical planetesimals, these parameters were assigned based on plausible values derived from planet formation theory. All parameter values used in the simulations are detailed in the Appendix~\ref{sec:appendix_traj}.

In addition to the gravitational force, the spacecraft experiences acceleration due to solar radiation pressure. To ensure safety during the simulation, the spacecraft is designed to perform an instantaneous avoidance maneuver by reversing its velocity upon reaching either a defined \emph{safety} ellipsoid or an \emph{exit} sphere. The safety ellipsoid is defined with axes equal to 1.4 times the corresponding asteroid dimensions, serving as a protective buffer zone around the asteroid. The exit sphere has a radius twice that of the asteroid's exterior Brillouin sphere, acting as a boundary beyond which the spacecraft exits the asteroid’s gravitational influence.

\begin{figure}
    \centering
    \includegraphics[width=1.0\linewidth]{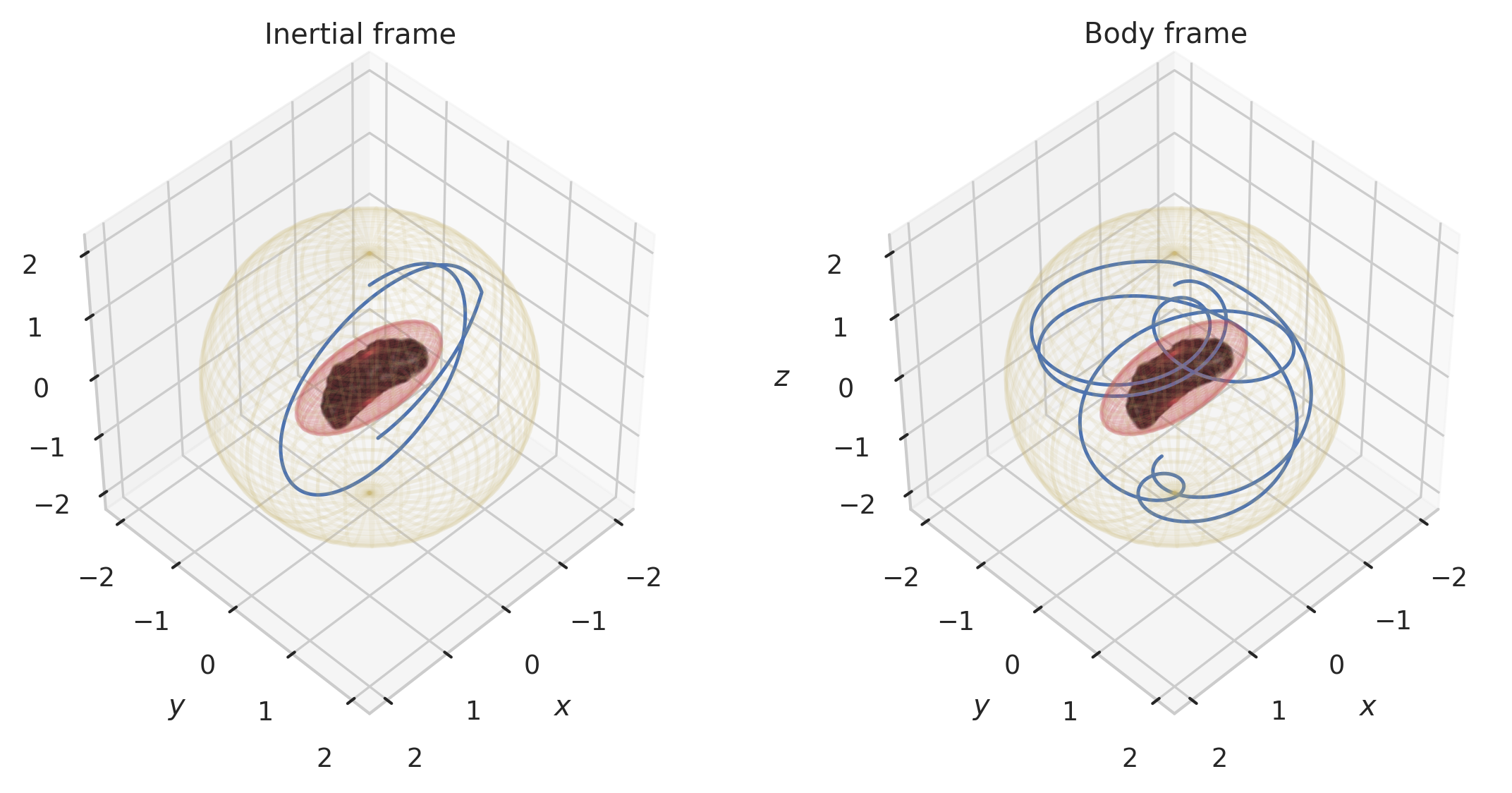}
    \caption{Trajectory of the spacecraft around the asteroid \emph{Eros – Three Regions} modeled with the ground-truth mascon gravity field. Trajectories are shown in both the inertial and body-fixed reference frames. The red ellipsoid represents the safety boundary, while the yellow sphere denotes the exit boundary.}
    \label{fig:traj_eros3}
\end{figure}

\begin{figure}
    \centering
    \imgrev{\includegraphics[width=1.0\linewidth]{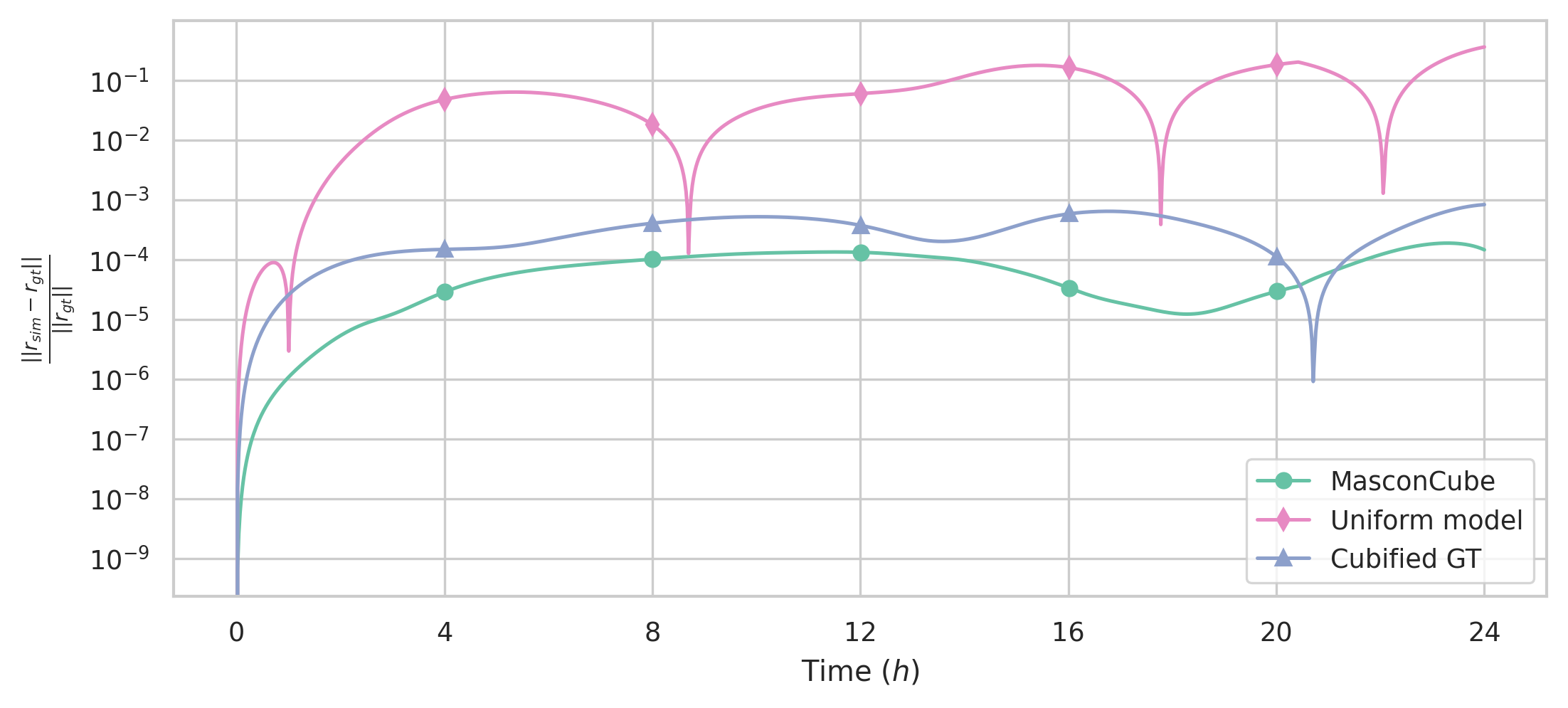}}
    \caption{Relative error of the spacecraft’s position vector at each timestep during the trajectory simulation around \textit{Eros – Three Regions}. This plot compares the accuracy of different gravity models against the ground-truth mascon model.}
    \label{fig:traj_eros3_err}
\end{figure}

Figure \ref{fig:traj_eros3} illustrates the resulting trajectory around the asteroid \textit{Eros – Three Regions} using the ground-truth mascon gravity model. We simulate spacecraft trajectories using four gravity representations: the ground-truth mascon model, its cubified approximation, the learned MasconCube model, and a uniform MasconCube. Figure \ref{fig:traj_eros3_err} presents the relative error of the spacecraft’s position vector $\mathbf{r}$ at each timestep for \textit{Eros – Three Regions}, while Table \ref{tab:traj_err} summarizes the relative position errors at the final timestep for all asteroids in the dataset.

\begin{table}
    \centering
    \caption{\rev{Trajectory simulation accuracy comparison: mean relative position errors for different gravity modeling approaches. Bold values indicate best performance for each asteroid.}}
    \begin{tabular}{lS[table-format=1.2e-1]S[table-format=1.2e-1]S[table-format=1.2e-1]}
    \toprule
        \multirow{2}{*}{\textbf{Asteroid}} & \multicolumn{3}{c}{\textbf{Mean relative error on $||\mathbf{r}||$}} \\
        & {\textbf{MasconCube}} & {\textbf{Uniform model}} & {\textbf{Cubified ground-truth}} \\
    \midrule
         Bennu & 3.86e-03 & 5.88e-02 & \textbf{\num{1.11e-03}} \\
         Eros - Two regions & \textbf{\num{3.31e-05}} & 1.75e-02 & 3.09e-04 \\
         Eros - Three regions & \textbf{\num{7.10e-05}} & 7.70e-02 & 3.12e-04 \\
         Eros - Uniform & 1.57e-05 & \textbf{\num{4.16e-15}} & 3.50e-04 \\
         Itokawa - Two regions & \textbf{\num{9.81e-06}} & 1.62e-01 & 9.57e-04 \\
         Itokawa - Smooth & \textbf{\num{2.31e-04}} & 6.59e-02 & 9.52e-04 \\
         Planetesimal - Hollow & \textbf{\num{2.51e-05}} & 1.29e-02 & 4.68e-04 \\
         Planetesimal - Hollow decentered & 3.33e-04 & 5.53e-02 & \textbf{\num{2.13e-04}} \\
         Planetesimal - Uniform & 6.27e-05 & \textbf{\num{4.26e-15}} & 6.68e-04 \\

    \bottomrule
    \end{tabular}
    \label{tab:traj_err}
\end{table}

\subsection{Training Time}

One of the most significant practical advantages of MasconCubes over competing approaches lies in their computational efficiency during training. MasconCubes demonstrate superior training efficiency compared to GeodesyNets by one order of magnitude, primarily due to fundamental algorithmic differences that distinguish mascon-based optimization from neural network approaches.
Unlike GeodesyNets which require computationally expensive numerical integration procedures to evaluate density distributions at arbitrary points within the body, in MasconCubes the gravitational acceleration prediction involves only straightforward vector arithmetic and norm calculations, eliminating the need for iterative numerical methods that characterize implicit neural representations.

They also outspeed other existing approaches in literature, as shown in Table~\ref{tab:speed}. Results are obtained on an NVIDIA GeForce RTX 2080 Ti with 11 \unit{GB} of VRAM \rev{with a a grid size equal to 100 for the MasconCube and equal to 40 for the MasconCube-S. All the models are trained using the Adam optimizer. As shown in Table~\ref{tab:speed_step}, training step time scales linearly with batch size, while grid density has negligible impact, thanks to efficient vectorized operations independent of grid resolution.}

\begin{table}
    \centering
\caption{\rev{Comparison of training time required to reach Table~\ref{tab:acc}'s accuracy on an NVIDIA RTX 2080 Ti.}}
    \begin{tabular}{lS[table-format=1.3]S[table-format=5]S[table-format=3.3]S[table-format=4]S[table-format=5]}
    \toprule
        \multirow{2}{*}{\textbf{Model}} & {\textbf{Training step time}} & {\textbf{Num. steps}} & {\textbf{Total training time}} & {\rev{\textbf{Batch size}}} & {\rev{\textbf{Parameters}}} \\
        & {(\unit{s})} & & {(\unit{min})} & & \\
    \midrule
    GeodesyNet & 1.75 & 10000 & 291.67 & 1000 & 80800 \\
    PINN-GM III & 1.90 & 8192 & 259.41 & 4096 & 2211 \\
    MasconCube & 0.396 & 1000 & 6.6 & 1000 & 35603 \\
    \rev{MasconCube-S} & 0.067 & 400 & 0.45 & 100 & 2160 \\
    \bottomrule
    \end{tabular}
    \label{tab:speed}
\end{table}

\begin{table}
    \centering
    \caption{\rev{Comparison of training step time with different settings on an NVIDIA RTX 2080 Ti.}}
    \rev{
    \begin{tabular}{lrcccc}
        \toprule
        & & \multicolumn{4}{c}{\textbf{Training step time (\unit{s})}} \\
        \midrule
        & & \multicolumn{4}{c}{\textbf{Grid size}} \\
        \cmidrule(lr){3-6}
        & & {25} & {50} & {100} & {200} \\
        \midrule
         \multirow{4}{*}{\textbf{Batch size}} & 250{\quad} & 0.109 & 0.112 & 0.112 & 0.111 \\
          & 500{\quad} & 0.218 & 0.218 & 0.215 & 0.211 \\
           & 1000{\quad} & 0.397 & 0.407  & 0.396 & 0.397 \\
           & 2000{\quad} & 0.761 & 0.750  & 0.766 & 0.766 \\
        \bottomrule
    \end{tabular}}
    \label{tab:speed_step}
\end{table}

The reduced computational requirements make MasconCubes particularly well-suited for scenarios requiring rapid model updates or real-time adaptation to new gravitational observations. 

\rev{While on-board training for spacecraft applications remains computationally challenging due to power and hardware constraints, MasconCubes represent a significant step toward this capability. Recent advances in space-grade neural network implementations on radiation-hardened processors demonstrate the feasibility of deploying machine learning algorithms on spacecraft for autonomous operations~\cite{mystkowska2025hardware, diana2024review}, including on-board gravity inversion~\cite{rudge2025memristor}. MasconCubes' simplified computational architecture, requiring only direct point-mass gravity calculations without iterative numerical integration, substantially reduces the gap between current space-based computing capabilities and on-board adaptive gravity modeling. Moreover, the structure of MasconCubes makes them amenable to implementation on specialized hardware accelerators that are commonly deployed in space applications for their radiation tolerance and power efficiency. Such implementations could enable future missions to perform autonomous gravity field refinement during proximity operations around small bodies, advancing the state-of-the-art in spacecraft autonomy.}

\section{Open Challenges}

Extensive experimentation with MasconCubes has identified several factors that affect their implementation and performance. This section discusses these considerations and their practical implications.

\subsection{Requirement for Prior Shape Knowledge}

In this work, we assumed the shape of the body to be necessary information for training a MasconCube. However, from a theoretical point of view, it is possible to perform training without this information, with the gradient causing mascons outside the asteroid to converge to zero, similar to what occurs with standard, non-differential training of GeodesyNets.

In this section, we demonstrate that such attempts failed in our experiments and attempt to explain why this configuration is more challenging for MasconCubes than for GeodesyNets.

\begin{figure}
    \centering
    \subfigure[Training points in {[0, 1]}]{
        \includegraphics[width=0.28\linewidth]{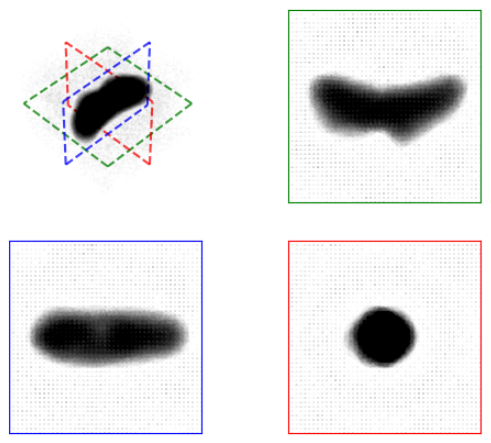}
        \label{fig:shape_free_training_1}
    }
    \hfill
    \subfigure[Training points in {[0.3, 1]}]{
        \includegraphics[width=0.28\linewidth]{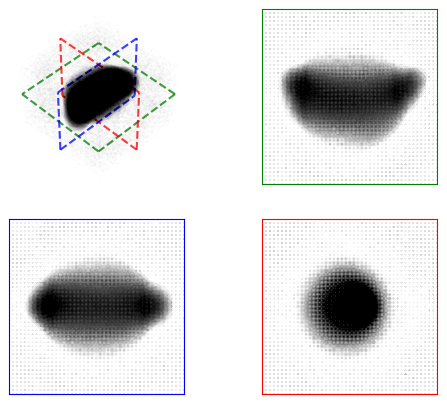}
        \label{fig:shape_free_training_2}
    }
    \hfill
    \subfigure[Training points in {[0.6, 1]}]{
        \includegraphics[width=0.28\linewidth]{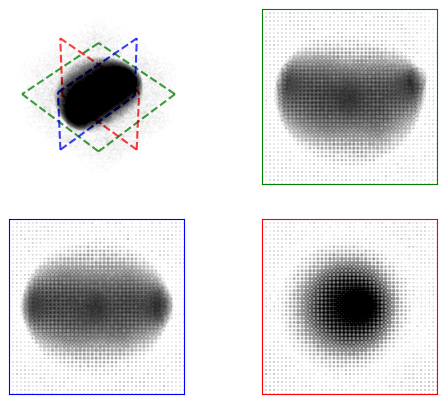}
        \label{fig:shape_free_training_3}
    }
    \caption{Shape-free MasconCube training results on Eros for different observation altitude ranges. Black points represent mascons with size proportional to mass. (a) Full altitude range [0, 1] correctly recovers asteroid shape. (b) Mid-range altitudes [0.3, 1] show partial degradation. (c) High altitudes only [0.6, 1] converge to spherical distribution. The progressive loss of shape fidelity demonstrates the challenge of gravity inversion from distant observations.}
    \label{fig:shape_free_training}
\end{figure}

Figure~\ref{fig:shape_free_training} shows the learned mass distribution of a MasconCube when the asteroid's shape is unknown, with training points sampled at different altitude ranges. When training points are sampled throughout the entire space around the asteroid, including close-to-surface points, the mass distribution, and consequently the shape, of the asteroid is correctly learned. However, as we begin sampling only points farther from the asteroid surface, the MasconCube training converges to spherical shapes in regions where no training points are sampled. This behavior is likely linked to the ill-posed nature of the gravity inversion problem, where multiple mass distributions can produce similar gravitational signatures at distant observation points, leading the optimization to favor simpler, more regular configurations in the absence of constraining data.

\begin{figure}
    \centering
    \subfigure[Gradient distribution using a minimal training batch of two observation points]{
        \includegraphics[width=0.45\linewidth]{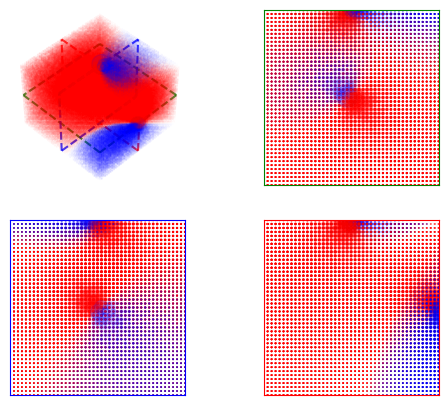}
        \label{fig:gradient_flow_comparison_1}
    }
    \hfill
    \subfigure[Gradient distribution using eight training points distributed on a circumference of radius 0.6]{
        \includegraphics[width=0.45\linewidth]{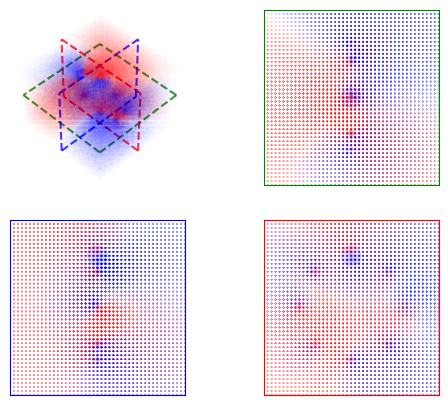}
        \label{fig:gradient_flow_comparison_2}
    }
   \caption{Gradient flow visualization in MasconCube training demonstrating the spatial localization effect. Red dots indicate regions where mascon masses should increase, blue dots indicate regions where masses should decrease, with dot size proportional to gradient magnitude. This spatial localization explains why MasconCubes struggle with shape-free training when observation points are distant from the asteroid surface, as only mascons near training points receive meaningful updates, leading to convergence toward spherical configurations in unobserved regions.}
    \label{fig:gradient_flow_comparison}
\end{figure}

The challenges are exacerbated by the explicit representation nature of MasconCubes compared to the implicit representation of GeodesyNets. This fundamental difference causes gradients to flow differently to parameters according to their spatial position, making it impossible for the model to properly learn the asteroid's shape when training points are located too far from the surface, as demonstrated in Figure~\ref{fig:gradient_flow_comparison}.

In explicit mascon representations, each grid point has a fixed spatial location, and gradients are computed based on the direct contribution of each mascon to the observed gravitational field. When observation points are distant from the asteroid, the gravitational signatures become increasingly ambiguous, and the optimization tends to distribute mass according to the simplest configuration that satisfies the constraints, typically a spherically symmetric distribution.

In contrast, implicit neural representations like GeodesyNets learn continuous density functions that can capture complex spatial relationships through their network architecture, allowing them to better extrapolate shape information even from distant observations.

Since we were able to successfully train MasconCubes without prior shape constraints only when dense surface observations were available during training, and given that such comprehensive surface coverage would already provide sufficient information to reasonably approximate the asteroid's shape, we concluded that explicit shape information represents a necessary prerequisite for our approach. Nevertheless, we believe that shape-free training, while challenging, should remain achievable, as demonstrated by the success of GeodesyNets in similar configurations.

Future work should explore proper regularization techniques that could enable shape-free training of MasconCubes. Potential approaches might include spatial regularization terms that encourage smooth mass distributions, or multi-scale training strategies.

This limitation represents an important trade-off in our current methodology: while requiring shape information constrains the applicability of MasconCubes to scenarios where prior shape knowledge is available, it significantly improves training stability and convergence reliability compared to unconstrained approaches.

\subsection{Hollow Body Modeling Challenges}
To evaluate the limits of our approach, we included two hollow body configurations: \textit{Planetesimal - Hollow} and \textit{Planetesimal - Hollow Decentered}. Previous literature has established that gravity inversion for hollow bodies represents a particularly challenging problem \cite{izzo2022geodesy}.

\begin{figure}
    \centering
    \subfigure[Ground Truth]{
        \includegraphics[width=0.31\linewidth]{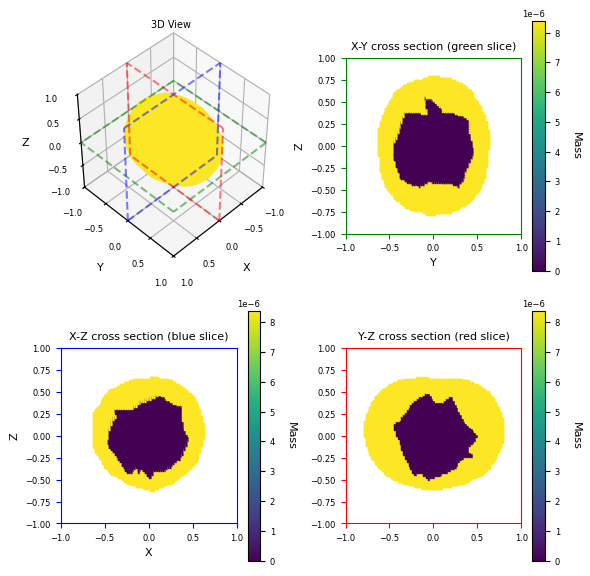}
        \label{fig:density_hollow_gt}
    }
    \hfill
    \subfigure[MasconCube]{
        \includegraphics[width=0.31\linewidth]{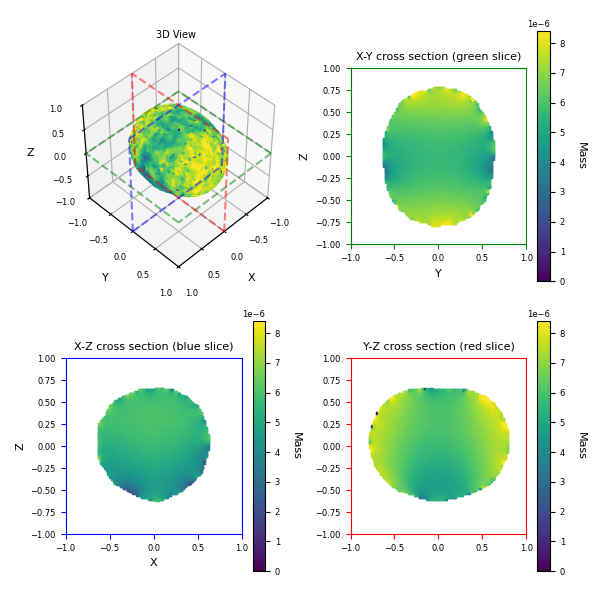}
        \label{fig:density_hollow_cube}
    }
    \hfill
    \subfigure[GeodesyNet]{
        \includegraphics[width=0.31\linewidth]{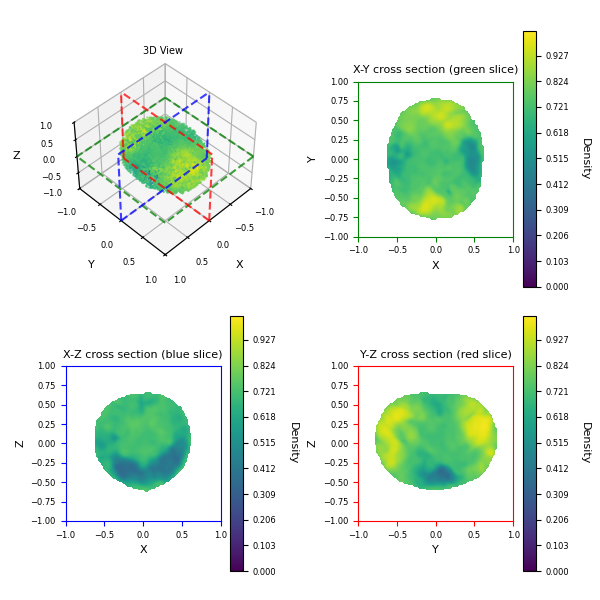}
        \label{fig:density_hollow_net}
    }
    \caption{Density reconstruction cross-sections for \emph{Planetesimal – Hollow} showing central cavity. Both methods fail to accurately reconstruct the internal void structure.}
    \label{fig:density_hollow}
\end{figure}

Hollow body gravity inversion is inherently difficult due to Newton's Shell theorem and the ill-posed nature of the inverse problem. According to Newton's Shell theorem, a spherically symmetric hollow shell produces the same external gravitational field as a point mass at its center, making internal cavities virtually undetectable from external observations alone \cite{newton1687principia}. This fundamental ambiguity is exacerbated for concentric configurations, where multiple density distributions can produce nearly identical gravitational signatures. The \textit{Planetesimal - Hollow} configuration features a central spherical cavity. As shown in Figure~\ref{fig:density_hollow}, MasconCube does not outperform GeodesyNet, with both methods struggling to reconstruct the internal void.

\begin{figure}
    \centering
    \subfigure[Ground Truth]{
        \includegraphics[width=0.31\linewidth]{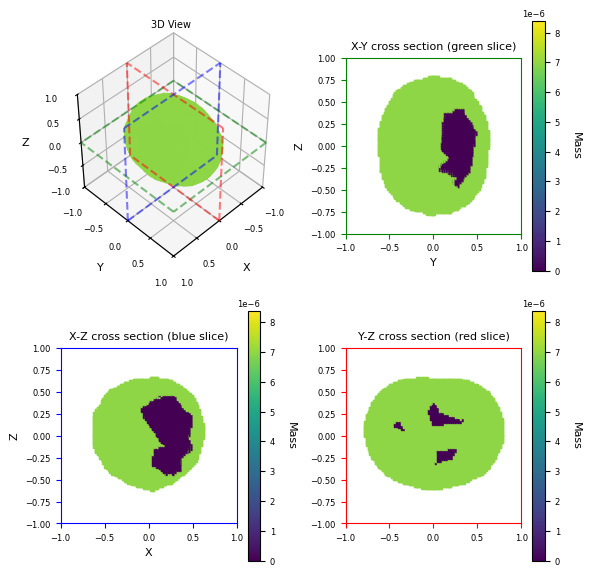}
        \label{fig:density_hollow_dec_gt}
    }
    \hfill
    \subfigure[MasconCube]{
        \includegraphics[width=0.31\linewidth]{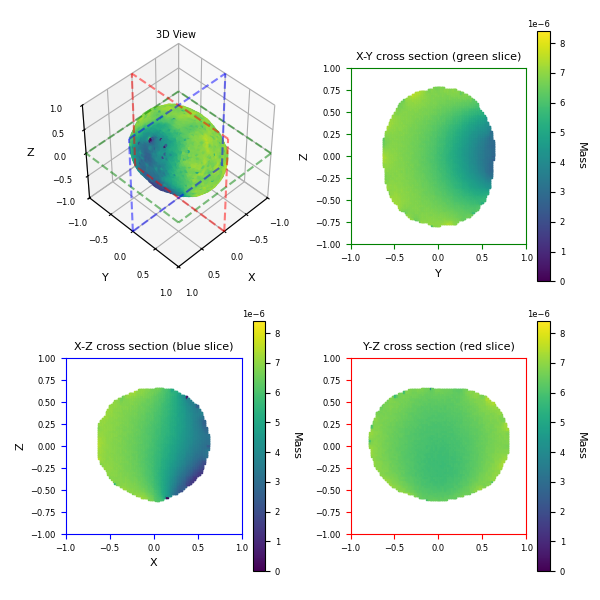}
        \label{fig:density_hollow_dec_cube}
    }
    \hfill
    \subfigure[GeodesyNet]{
        \includegraphics[width=0.31\linewidth]{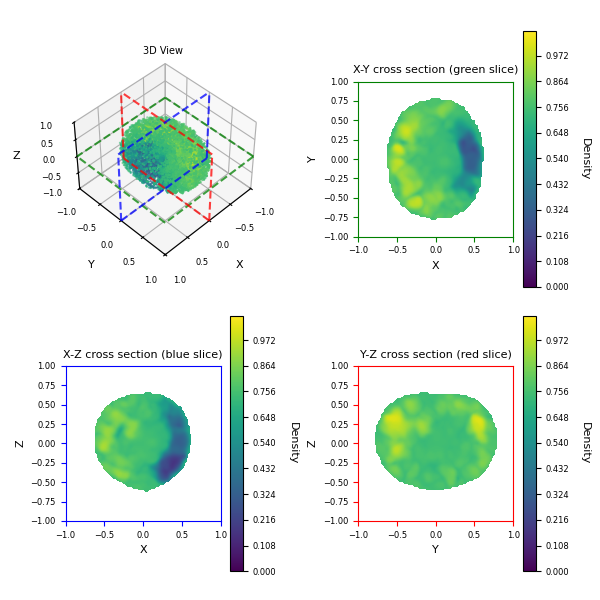}
        \label{fig:density_hollow_dec_net}
    }
    \caption{Density reconstruction cross-sections for \emph{Planetesimal – Hollow Decentered} with off-center cavity. Methods detect asymmetry but struggle with void reconstruction.}
    \label{fig:density_hollow_dec}
\end{figure}

We also tested an asymmetrical variant with a displaced hollow region, representing a theoretically less ambiguous case. As shown in Figure~\ref{fig:density_hollow_dec}, while both methods successfully detect the asymmetry, neither adequately reconstructs the hollow region itself. MasconCube shows particularly limited capability in resolving internal void structures, suggesting that explicit mascon representations require additional constraints for complex internal geometries.

These results demonstrate a fundamental limitation in gravity-based density reconstruction: external gravitational observations effectively constrain surface density variations but provide insufficient information about internal cavities, especially symmetric ones. \rev{This limitation also has broader implications for the study of asteroid porosity. Since macroporosity and distributed voids influence the bulk density in ways that can be gravitationally degenerate with uniform-density models, accurate porosity estimation requires integrating gravitational data with complementary observations. Our findings thus emphasize that purely gravity-based inversions can reveal surface and bulk mass anomalies but remain insensitive to internal porosity distributions, an important consideration for planetary modeling and mission data interpretation~\cite{baer2011astrometric, britt2001modeling, okada2020highly}.}

\section{Conclusions}

This research presents MasconCubes as a significant advancement in gravitational field modeling for irregularly shaped small bodies, successfully addressing the computational and accuracy limitations that have constrained existing methodologies. Through comprehensive evaluation across diverse asteroid configurations including Bennu, Eros, Itokawa, and synthetic planetesimals, MasconCubes demonstrate superior performance compared to established ML approaches such as GeodesyNets and PINN-GM III in modelling the internal density distribution of the bodies.

The self-supervised learning approach achieves remarkable improvements in modeling fidelity, with Mean Absolute Errors of normalized Stokes coefficients reduced by 1-2 orders of magnitude compared to GeodesyNets, and gravitational acceleration prediction errors decreased by nearly an order of magnitude. Trajectory simulation results further validate the practical utility of the approach. Most notably, MasconCubes achieve these performance gains while demonstrating computational efficiency approximately 40 times greater than competing neural network approaches, with training times reduced from hundreds of minutes to under 7 minutes.

The explicit physical interpretation provided by direct mass distribution modeling represents a key advantage over implicit neural representations, offering mission planners and scientists intuitive understanding of gravitational field characteristics and internal asteroid structure. The formulation of gravity inversion as direct optimization over a regular 3D mascon grid, combined with the normalized L1 loss function and optimal scaling factor, provides a robust framework that balances computational tractability with physical accuracy.

While the current methodology requires prior shape knowledge, which constrains applicability to well-characterized bodies, this limitation significantly improves training stability and convergence reliability. Shape-free training remains challenging due to the ill-posed nature of gravity inversion from distant observations, though future regularization techniques may address this constraint.

As deep space exploration missions increasingly target small bodies for scientific exploration, resource utilization, and planetary defense, MasconCubes provide the gravitational modeling foundation essential for mission success. The combination of high accuracy, computational efficiency, and physical interpretability positions this approach as a key enabling technology for the next generation of asteroid exploration missions, offering practical solutions for proximity operations while advancing our understanding of small body internal structure and composition.

\section*{References}
%% --- for bib
\bibliographystyle{astrobib}
\bibliography{refs}

\appendix
\section{Appendix}
\label{sec:appendix}
In this appendix we provide further experimental details and parameter specifications used throughout this study.

\subsection{Tetrahedral Mesh Models}
\label{sec:appendix_tetra}
Table~\ref{tab:meshes} summarizes the discretized tetrahedral mesh models used in this study, while Table~\ref{tab:density_functions} details the mathematical functions used to create the various density distributions for each asteroid configuration.

\begin{table}[H]
    \centering
    \caption{\rev{Summary of the tetrahedral mesh models. Volume values are dimensionless, as meshes are scaled.}}
    \begin{tabular}{l S[table-format=4.0] S[table-format=4.0] S[table-format=5.0] S[table-format=2.5]}

        \toprule
        \textbf{Body} & \textbf{Mesh Vertices} & \textbf{Mesh Triangles} & \textbf{Tetrahedra} & \textbf{Total Volume} \\
        \midrule
        Bennu        & 739 & 1474 & 6281   & 1.40759 \\
        Eros         & 739 & 1474 & 8094 & 0.29358 \\
        Itokawa      & 813 & 1622 & 8849 & 0.41142 \\
        Planetesimal & 1119 & 2234 & 11735 & 1.36642 \\
        \bottomrule
    \end{tabular}
    \label{tab:meshes}
\end{table}

\begin{table}[H]
    \centering
    
    \caption{\rev{Mass functions \(m(x,y,z,V)\) used to generate the models. Each mascon is placed at the centroid \((x,y,z)\) of a tetrahedral element with volume \(V\), and masses are normalized so the total body mass equals 1.}}

    \begin{tabular}{llc}

        \toprule
        \textbf{Mesh} & \textbf{Density distribution} & \textbf{Mass function $m(x,y,z,V)$} \\
        \midrule
        Bennu        & Three regions & $V\;\;\text{if}\; z\in[-0.3,0.3]\,\text{,}\;\;\;3V\;\text{otherwise}$  \\
        \arrayrulecolor{black!30}\midrule
        \multirow{3}{*}{Eros}   & Two regions & $1.5V\;\;\text{if}\; y<-0.1\,\text{,}\;\;\;V\;\text{otherwise}$  \\
                                & Three regions & $1.5V\;\;\text{if}\; x<-0.3\,\text{,}\;\;\; 0.5V\;\;\text{if}\;x-\frac{z}{2}<-0.3\,\text{,}\;\;\;V\;\text{otherwise}$ \\
                                & Uniform & $V$ \\
        \arrayrulecolor{black!30}\midrule
        \multirow{2}{*}{Itokawa}    & Two regions  & $\frac{57}{35}V\;\;\text{if}\; x-\frac{z}{2}<-0.3\,\text{,}\;\;\;V\;\text{otherwise}$  \\
                                    & Smooth & $V\cos(1.5x)$ \\
        \arrayrulecolor{black!30}\midrule
        \multirow{3}{*}{Planetesimal}   & Hollow & $0\;\;\text{if}\; x^2+y^2+z^2<0.2\,\text{,}\;\;\;V\;\text{otherwise}$ \\
                                        & Hollow decentered & $0\;\;\text{if}\; 4(x-0.25)^2+y^2+z^2<0.2\,\text{,}\;\;\;V\;\text{otherwise}$ \\
                                        & Uniform & $V$ \\
        \arrayrulecolor{black}\bottomrule
    \end{tabular}
    \label{tab:density_functions}
\end{table}

\subsection{Training Parameters}
\label{sec:appendix_training}

Table~\ref{tab:training} provides the complete set of hyperparameters used for MasconCube training across all experiments presented in this work.

\begin{table}[H]
\centering
\caption{\rev{Complete hyperparameter settings for MasconCube training.}}
\label{tab:hyperparameters}
\rev{
\begin{tabular}{ll}
\toprule
\textbf{Hyperparameter} & \textbf{Value} \\
\midrule
\multicolumn{2}{l}{\textit{Grid Configuration}} \\
Grid resolution & $100 \times 100 \times 100$ \\
Number of active mascons ($\hat{N}$) & $< 10^6$ (body-dependent) \\
\midrule
\multicolumn{2}{l}{\textit{Initialization}} \\
Mass initialization distribution & $U(-\frac{1}{10\hat{N}}, \frac{1}{10\hat{N}})$ \\
Position initialization & Fixed grid points \\
Random seed & 40 \\
\midrule
\multicolumn{2}{l}{\textit{Training Configuration}} \\
Total training steps & 1000 \\
Batch size & 1000 \\
Batch refresh frequency & Every 10 epochs \\
Sampling region & Sphere of radius 1.0 \\
\midrule
\multicolumn{2}{l}{\textit{Optimizer Settings}} \\
Optimizer & Adam \cite{kingma2017adam} \\
Initial learning rate & $1 \times 10^{-5}$ \\
Learning rate schedule & Step decay \\
Step decay factor & 0.8 \\
Step decay intervals & 200 \\
$\beta_1$ (Adam) & 0.9 (default) \\
$\beta_2$ (Adam) & 0.999 (default) \\
$\epsilon$ (Adam) & $1 \times 10^{-8}$ (default) \\
Weight decay & 0 \\
\midrule
\multicolumn{2}{l}{\textit{Loss Function}} \\
Loss type & Normalized L1 (Equation~\ref{eq:loss}) \\
Scaling factor computation & Closed-form optimal $c$ \\
\midrule
\multicolumn{2}{l}{\textit{Regularization}} \\
Mass normalization constraint & $\sum_j m_j = 1$ (enforced) \\
\bottomrule
\end{tabular}
\label{tab:training}
}
\end{table}

\subsection{Trajectory Integration Parameters}
\label{sec:appendix_traj}
Table~\ref{tab:trajectory} lists the physical parameters used in the spacecraft trajectory simulations, including gravitational and orbital characteristics required for realistic dynamical modeling.

\begin{table}[H]
    \centering
    \caption{\rev{Physical parameters used in spacecraft trajectory simulations. Values for Itokawa, Bennu, and Eros are from observations, while planetesimal parameters are theoretical, derived from planet formation models.}}
    \begin{tabular}{lS[table-format=1.3e2]S[table-format=2.2]S[table-format=1.2]S[table-format=1.2]S[table-format=5.0]}
        \toprule
        \textbf{Asteroid} & {\textbf{Mass}} & {\textbf{Rotation period}} & {\textbf{Aphelion distance}} & {\textbf{Perihelion distance}} & {\textbf{Longest dimension}} \\
        & {(\unit{kg})} & {(\unit{h})} & {(\unit{AU})} & {(\unit{AU})} & {(\unit{m})} \\
        \midrule
        Itokawa & 3.51e10 & 12.1 & 1.69 & 0.95 & 535 \\
        Bennu & 7.329e10 & 4.3 & 1.35 & 0.89 & 565 \\
        Eros & 6.69e15 & 5.27 & 1.78 & 1.13 & 34400 \\
        Planetesimal & 1.0e11 & 10 & 1.5 & 1.0 & 1000 \\
        \bottomrule
    \end{tabular}
    \label{tab:trajectory}
\end{table}

\clearpage
\subsection*{Authors Biography}
% \note{(at least the first author's and the corresponding author's)}

\begin{biography}[pietro]{Pietro Fanti} holds a Master’s degree in Artificial Intelligence from the University of Bologna (Italy) and a Bachelor's degree in Computer Science from the University of Florence. He currently works at the European Space Agency’s Advanced Concepts Team. His research integrates AI techniques with space sciences to address complex problems in gravitational modeling and spacecraft dynamics. Prior to this, Pietro gained experience in applied AI research in industry, focusing on generative computer vision and developing scalable AI solutions for widely used applications.
\end{biography}

\begin{biography}[dario]{Dario Izzo} graduated as a Doctor of Aeronautical Engineering from the University Sapienza of Rome (Italy). He then took a second master in Satellite Platforms at the University of Cranfield in the United Kingdom and completed his Ph.D. in Mathematical Modelling at the University Sapienza of Rome where he lectured classical mechanics and space flight mechanics. Dario Izzo later joined the European Space Agency and became the scientific coordinator of its Advanced Concepts Team. He devised and managed the Global Trajectory Optimization Competitions events and the Kelvins innovation and competition platform and made key contributions to the understanding of flight mechanics and spacecraft control and pioneering techniques based on evolutionary and machine-learning approaches. Dario Izzo received the Humies Gold Medal, the Barry Carlton award and led the team winning the 8th edition of the Global Trajectory Optimization Competition.
\end{biography}

% \begin{biography}[<author figure name>]{Author Name}
% introduction
% \end{biography}

\vspace*{2.6em}
\subsection*{Graphical table of contents}

\begin{figure}[H]
    \centering
    \includegraphics[width=1.0\linewidth]{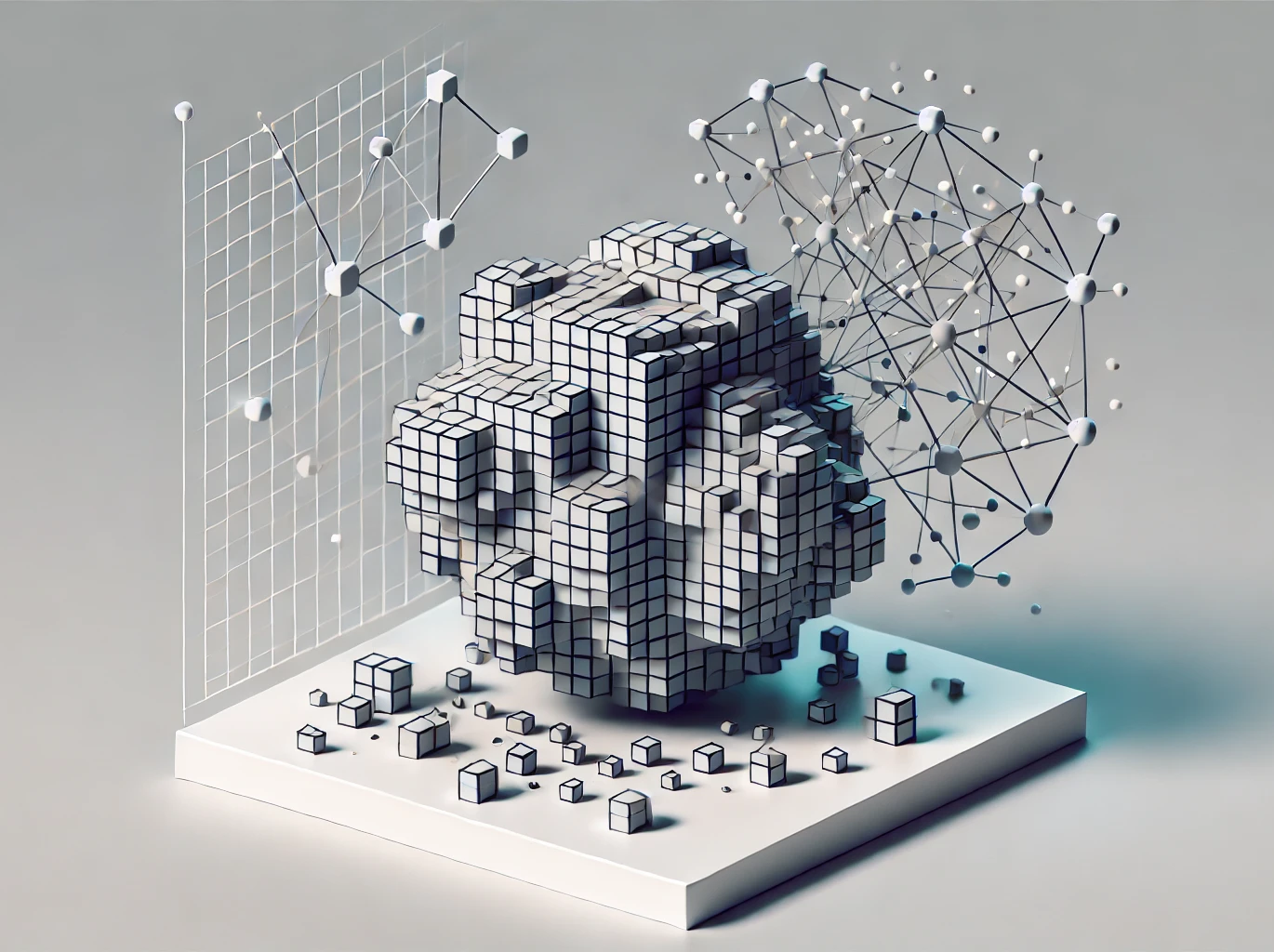}
    \caption*{MasconCubes, a self-supervised learning framework, addresses the challenges of gravitational modeling for irregular small bodies by directly optimizing mass distributions on a 3D mascon grid. It achieves higher accuracy, 40× faster training, and improved physical interpretability compared to other neural approaches, making it highly suitable for asteroid and comet exploration.}
\end{figure}

\end{document}